\PassOptionsToPackage{noend}{algorithmic}
\PassOptionsToPackage{dvipsnames}{xcolor}
\PassOptionsToPackage{breakspaces}{clevethm}
\PassOptionsToPackage{hypertexnames=false}{hyperref}
\PassOptionsToPackage{capitalize,nameinlink}{cleveref}
\PassOptionsToPackage{textsize=footnotesize}{todonotes}

\documentclass[journal,10pt,twoside]{IEEEtranTCOM}
\usepackage{graphics}
\usepackage{adjustbox}
\usepackage{pgfplots}
\usepackage[outdir=./TeX/images/]{epstopdf}
\usepackage{caption}
\usepackage{subcaption}
\usepackage{xspace}
\usepackage{cite}
\usepackage[normalem]{ulem}
\usepackage{url}

\normalsize
\usepackage{booktabs}
\usepackage{makecell}
\usepackage{float}
\usepackage{algorithmic}
\usepackage{bbm}
\usepackage{letltxmacro}
\usepackage[most]{tcolorbox}
\usepackage[%
	thmreset=section,
	eqreset=section,
	noload={algpseudocode,algorithmic}
]{myPreamble}

\pgfplotsset{compat=1.16}
\allowdisplaybreaks

\newlength\myindent
\newcommand{\sqnrm}[1]{\left\Vert{#1}\right\Vert^2}
\newcommand{\nrm}[1]{\left\Vert{#1}\right\Vert}

\DeclareMathOperator\vect{vec}
\DeclareMathOperator\filt{filtration}

\makeatletter
	\newcommand{\E}{\@ifstar\@@E\@E}					% Conditional expectation \E[k]{arg}[conditional event] (use *-version for small brackets)
	\newcommandx{\@E}[3][1={},3={}]{
	\mathbb E_{#1}\ifstrempty{#2}{}{
		\left[
			#2\ifstrempty{#3}{}{\mid{#3}}
		\right]
	}
	}
	\newcommandx{\@@E}[3][1=k,3={}]{
	\mathbb E_{#1}\ifstrempty{#2}{}{
		[#2\ifstrempty{#3}{}{|#3}]
	}
	}

	\newcommand{\hatE}{\@ifstar\@@hatE\@hatE}					% Conditional expectation \E[k]{arg}[conditional event] (use *-version for small brackets)
	\newcommandx{\@hatE}[3][1={},3={}]{
	\hat{\mathbb E}_{#1}\ifstrempty{#2}{}{
		\left[
			#2\ifstrempty{#3}{}{\mid{#3}}
		\right]
	}
	}
	\newcommandx{\@@hatE}[3][1=k,3={}]{
	\hat{\mathbb E}_{#1}\ifstrempty{#2}{}{
		[#2\ifstrempty{#3}{}{|#3}]
	}
	}

\newcommand{\subalign}[1]{%
  \vcenter{%
    \Let@ \restore@math@cr \default@tag
    \baselineskip\fontdimen10 \scriptfont\tw@
    \advance\baselineskip\fontdimen12 \scriptfont\tw@
    \lineskip\thr@@\fontdimen8 \scriptfont\thr@@
    \lineskiplimit\lineskip
    \ialign{\hfil$\m@th\scriptstyle##$&$\m@th\scriptstyle{}##$\hfil\crcr
      #1\crcr
    }%
  }%
}
\makeatother

\Crefname{ALC@unique}{Line}{Lines}
\renewcommand{\algfont}{\bf}

\newlist{algsubstates}{enumerate}{2}
\makeatletter
	\setlist[algsubstates,1]{
		label={\alph*:},
		ref={\theALC@line.\alph*},
		itemsep=0pt,
		partopsep=0pt,
		topsep=0pt,
		parsep=0pt,
	}
\makeatother
\newcommand{\fixstate}{\addtocounter{ALC@line}{-1}\refstepcounter{ALC@line}\phantomsection}
\newcommand{\State}{\STATE\fixstate}

\makeatletter
	\renewcommand\theALC@line{\oldstylenums\arabic{ALC@line}}
	\def\theHALC@line{\thealgorithm-\arabic{ALC@line}}
\makeatother
\crefname{ALC@line}{step}{steps}
\Crefname{ALC@line}{Step}{Steps}
\crefalias{algsubstatesi}{ALC@line}

\newlist{tblsubstates}{enumerate}{2}
\makeatletter
	\setlist[tblsubstates,1]{
		label={\alph*:},
		ref={\theALC@line.\alph*},
		itemsep=0pt,
		partopsep=0pt,
		topsep=0pt,
		parsep=0pt,
	}
\makeatother
\crefname{TBL@line}{table}{tabless}
\Crefname{TBL@line}{table}{Tabless}
\crefalias{tblsubstatesi}{TBL@line}

\newenvironment{talign*}
 {\let\displaystyle\textstyle\csname align*\endcsname}
 {\endalign}

  \newcommand{\Affiliation}{%
  Pourya Behmandpoor, Panagiotis Patrinos, and Marc Moonen are with KU Leuven University, Department of Electrical Engineering (ESAT), STADIUS Center for Dynamical Systems, Signal Processing and Data Analytics (e-mail: pourya.behmandpoor, marc.moonen, panos.patrinos @esat.kuleuven.be).
  }
 
  \newcommand{\Thanks}{This research work was carried out at the ESAT Laboratory of KU Leuven, in the frame of Research Project FWO nr. G0C0623N 'User-centric distributed signal processing algorithms for next generation cell-free massive MIMO based wireless communication networks' and Fonds de la Recherche Scientifique - FNRS and Fonds voor Wetenschappelijk Onderzoek - Vlaanderen EOS Project no 30452698 '(MUSE-WINET) MUlti-SErvice WIreless NETworks'. 
  The Work is also supported by the Research Foundation Flanders (FWO) research projects G081222N, G033822N, and G0A0920N; Research Council KU Leuven C1 project No. C14/24/103;
  The scientific responsibility is assumed by its authors.}
\tikzexternaldisable

\title{Asynchronous Message-Passing and Zeroth-Order Optimization Based Distributed Learning with a Use-Case in Resource Allocation in Communication Networks}
\author{Pourya Behmandpoor, 
Marc Moonen,
Panagiotis Patrinos
\thanks{\Thanks \\\Affiliation}
}

\begin{document}
\maketitle

    \begin{abstract}
    Distributed learning and adaptation have received significant interest and found wide-ranging applications in machine learning and signal processing. While various approaches, such as shared-memory optimization, multi-task learning, and consensus-based learning (\eg federated learning and learning over graphs), focus on optimizing either local costs or a global cost, there remains a need for further exploration of their interconnections. This paper specifically focuses on a scenario where agents collaborate towards a common task (\ie optimizing a global cost equal to aggregated local costs) while effectively having distinct individual tasks (\ie optimizing individual local parameters in a local cost). Each agent's actions can potentially impact other agents' performance through interactions. Notably, each agent has access to only its local zeroth-order oracle (\ie cost function value) and shares scalar values, rather than gradient vectors, with other agents, leading to communication bandwidth efficiency and agent privacy. Agents employ zeroth-order optimization to update their parameters, and the asynchronous message-passing between them is subject to bounded but possibly random communication delays. This paper presents theoretical convergence analyses and establishes a convergence rate for nonconvex problems. Furthermore, it addresses the relevant use-case of deep learning-based resource allocation in communication networks and conducts numerical experiments in which agents, acting as transmitters, collaboratively train their individual policies to maximize a global reward, \eg a sum of data rates.
\end{abstract}

    \section{Introduction}
        
\subsection{Distributed learning over message-passing architectures}\label{sec:intro:A}
The recent proliferation of edge devices with computational power, along with increased data generation and processing, has motivated research on distributed learning and adaptation.  
In distributed learning, edge devices, as distributed agents, aim to cooperatively learn a common task or multiple individual tasks. 
This collaboration occurs within a message-passing architecture, which poses challenges due to heterogeneity in local cost functions/data, processing capabilities, and potential communication delays or failures.

Asynchronous distributed optimization and learning techniques in the literature address this collaboration and consider various message-passing approaches as follows:
\emph{Shared-memory approach:} All agents have access to a common shared memory \cite{cannelli_asynchronous_2018,peng_arock_2016,liu_asynchronous_2015,latafat_primal-dual_2022,bertsekas_parallel_2015}. Each agent reads the required information from the shared memory and updates the decision variable, either in full or in part, by writing to the shared memory. This model operates asynchronously as the reading and writing processes occur asynchronously.
\emph{Local-memory approach:} Each agent possesses a local (private) memory and asynchronously updates a specific part of the decision variable, \ie the local parameters \cite{bertsekas_parallel_2015,zhou_distributed_nodate,latafat_primal-dual_2022,latafat_new_2019}. After each update, the agent may broadcast the updated local parameter to other users who require it for their own updates. 
Due to random communication delays or dropouts, the information received from other agents and stored in the local memories may be outdated. 
The local-memory approach can also be considered, for instance, in multitask learning approaches \cite{nassif2020multitask} where each agent optimizes its local parameters based on the sum of its local cost function and a regularizer that captures the relation of the local parameters with other agents' local parameters. 
In games and multi-agent reinforcement learning, this approach can be considered where each agent optimizes its local cost function while capturing its dependence on other players through interaction \cite{narang_learning_2022,yang2020overview}.
\emph{Consensus-based approach:} Each agent updates its local parameters (asynchronously).
However, unlike the previous approaches, agents in this approach aim to achieve parameter consensus. 
Consensus can be achieved through frequent averaging facilitated by a central entity, as seen in federated learning settings \cite{mcmahan_communication-efficient_2017,li_federated_2020,yu_parallel_2019}, or through frequent averaging of local parameters among neighboring agents, as seen in learning over graph settings \cite{vlaski_distributed_2021-1,nassif_learning_2020,sayed_adaptation_2014,wu_decentralized_2017,duchi_dual_2012,hajinezhad2019zone}. 
In this approach, all agents strive to reach common optimal parameters optimized over a global cost function, such as the sum of local costs.

While the three aforementioned approaches in distributed optimization and learning have been extensively studied in isolation, their interconnections remain relatively unexplored. This motivates us to further investigate cases where agents share a common task while they also have different individual tasks. This scenario bridges the gap between the last two mentioned approaches, namely the local-memory and consensus-based approaches.

In this setting, each agent's actions and decisions can potentially impact the performance of other agents, necessitating consideration of interactions during local optimization, similar to the local-memory approach. Furthermore, agents optimize their local parameters not only based on their own local costs but also with respect to a global function (\eg sum of local costs) that reflects the common task, similar to the consensus-based approach.

To the best of our knowledge, this is the first theoretical convergence study addressing such a scenario, where agents asynchronously update their local parameters using zeroth-order oracles by querying only local cost function values. 
Employing zeroth-order optimization with zeroth-order oracles rather than first-order oracles (\ie function gradients) in each agent brings multiple benefits. 
While first- or higher-order oracles may not be available or expensive to calculate in some settings (\eg simulation-based optimization or bandit optimization \cite{hajinezhad2019zone}), zeroth-order oracles can be available in a more computationally- and memory-efficient manner \cite{zhang2024revisiting}. Moreover, zeroth-order oracles significantly reduce the communication bandwidth requirements in the proposed distributed learning as it helps agents share only scalars. This feature, sharing scalars unlike sharing local parameters or gradients in federated learning or learning over graph settings, ensures privacy-preserving cooperation as well \cite{zhang2022robustify} (refer to \cref{sec:proposed_alg} for more details).
The relevant problem of resource allocation (RA) has been examined as a use-case within this framework.

\subsection{Resource allocation in communication networks} \label{sec:intro:B}
RA has been an active research area in both wired and wireless communication networks for several decades. %\cite{cendrillon_autonomous_2007,luo_dynamic_2008}. 
Recently, deep learning (DL)-based approaches have emerged as a promising solution, offering improvements in speed, performance efficiency, and implementation simplicity compared to conventional RA methods \cite{zhou_dynamic_2019,tong_nine_2022}.

DL-based RA typically consists of two stages: \emph{the training stage} and \emph{the inference stage}. 
In the training stage, policy (typically a deep neural network (DNN)) parameters are optimized using data-driven approaches. In the inference stage, the trained policies are utilized for performing RA. 
Both training and inference stages can be categorized into two main scenarios: \emph{(i) centralized, (ii) distributed.} 

In the centralized inference scenario, a centralized policy implemented on a server performs the RA by collecting all the relevant global measurements  \cite{kalogerias_model-free_2020,liang_towards_2020,behmandpoor_deep_2021,behmandpoor_learning_2022,dong_deep_2021}. 
On the other hand, in the distributed scenario, each transmitter possesses its own (possibly unique \cite{liang_spectrum_2019}) RA policy, where it allocates resources locally, potentially through message passing with other transmitters \cite{liang_spectrum_2019,ye_deep_2019,nasir_multi-agent_2019,wang_learning_2022,naderializadeh_learning_2022,eisen_optimal_2020,naderializadeh_state-augmented_2022,shen_graph_2021,behmandpoor_federated_2022}. 
In both scenarios, however, the training stage is typically performed in a centralized manner \cite{ye_deep_2019,nasir_multi-agent_2019,liang_spectrum_2019,wang_learning_2022,naderializadeh_learning_2022,eisen_optimal_2020,naderializadeh_state-augmented_2022,dong_deep_2021}. %+\cite{kalogerias_model-free_2020,}

Distributed training facilitates online training of RA policies meaning that the policies can be trained simultaneously while they are being utilized. This offers several advantages over centralized (offline) training:
(1) Online training enables the capture of real-time channel and user state distributions, resulting in more accurate policy training compared to offline training, which may suffer from a mismatch between the training sample distribution and the actual sample distribution \cite{kalogerias_model-free_2020}.
(2) Non-stationary channels can lead to distribution drifts, which are caused by the dynamic nature of communication systems. 
Factors such as transmitter and receiver dynamic locations, varying environmental conditions (\eg, varying traffic patterns in vehicle-to-vehicle communication), and shifting user demand patterns can contribute to distribution drifts \cite{eisen_learning_2018,dong_deep_2021}. 
Distributed training provides a means to address these distribution mismatches and drifts through (online) policy retraining \cite{dong_deep_2021}.
(3) In practical scenarios, it may be infeasible to have a server with sufficient communication and computational capacity to perform centralized training. Distributed training offers a viable alternative in such cases, as it does not rely on a central server and can distribute the training process across multiple agents \cite{behmandpoor_MODEL-FREE_2022}. 

Online (distributed) training can be established by employing zeroth-order or higher-order optimization approaches.
Zeroth-order optimization for RA policy training (also called model-free training) is considered in this paper. During the proposed zeroth-order training, agents approximate utility function gradients by \emph{measuring} their utility values, \eg data rates, rather than calculating function gradients \cite{kalogerias_model-free_2020}. 
This approach can benefit from the advantages of zeroth-order optimization mentioned in the previous subsection. Moreover, it captures the full behavior of the communication system, including nonidealities such as nonlinearities in the modulator and demodulator or antenna setup, which are not typically captured by higher-order approaches due to assumed simplifications in the utility functions, \eg the simplifications made in data rate formulation.
The contributions are listed as follows:
\begin{enumerate}
    \item A novel distributed learning approach is proposed, where agents share a common task (hence they optimize the sum of local costs), while each agent has also an individual task (hence agents have local (unique) parameters).
    Agents impact each others' performance through interactions. They update their local parameters using zeroth-order optimization methods.
    During the training, agents share scalar quantities, which saves communication bandwidth and preserves their privacy. 
    \item The proposed learning approach operates asynchronously, with agents updating and querying their zeroth-order oracle at the time of interest. Communication between agents is subject to bounded but potentially random delays.
    \item Theoretical convergence analyses for smooth nonconvex functions are provided, establishing a convergence rate of $\mathcal{O}(\nicefrac{1}{\sqrt{t}})$, where $t$ represents the iteration counter. This convergence rate aligns with the state-of-the-art results found in the zeroth-order optimization literature.
    \item The relevant problem of DL-based RA is also explored. Building upon the proposed learning approach, agents, acting as transmitters, collaborate to achieve maximum global reward. The cross-talk communication channels naturally link the agents together.
\end{enumerate}
    \section{System Model}\label{sec:system_model}
        
In this paper, we examine the following optimization problem:
\begin{align}
\minimize_{\bm{\theta}=(\bm\theta_1,\cdots,\bm\theta_m)\in\R^n}{f(\bm{\theta}) \coloneqq \frac{1}{m}\sum_{\ell=1}^m f_\ell(\bm{\theta})},
\label{eq:opt}
\end{align}
where the functions $\func{f_\ell}{\R^n}{\R}$ are continuously differentiable but possibly nonconvex. We consider a scenario with $m$ agents aiming to cooperatively optimize \eqref{eq:opt} over a message-passing architecture. 
Each agent $\ell$ optimizes its local parameter vector $\bm\theta_\ell \in \R^{n_\ell}$, while its local cost function $f_\ell$ depends on $\bm{\theta} = (\bm\theta_\ell, \bm\theta_{-\ell})$, and $\bm\theta_{-\ell}\coloneqq \{\bm\theta_1,...,\bm\theta_{\ell-1},\bm\theta_{\ell+1},...,\bm\theta_m\}$ represents local parameters of other agents.
Furthermore, motivated by the advantages of zeroth-order optimization mentioned in the previous section, we consider the case where agents access only local stochastic zeroth-order oracles. In other words, each agent $\ell$ can only measure its noisy cost function value $F_\ell(\bm{\theta}, \bm{\xi})$ which is unbiased, \ie $\E[\bm{\xi} \sim \mathcal{D}^{\xi}_\ell]{F_\ell(\bm{\theta}, \bm{\xi})} = f_\ell(\bm{\theta})$. In this oracle, $\bm{\xi} \in \R^s$ is a random sample vector drawn from local data distribution $\mathcal{D}^{\xi}_\ell$ and may represent training samples in the learning context (cf. the example of RA in the use-case discussed in \cref{sec:RA}).
We consider the following basic assumptions for \eqref{eq:opt}:
\begin{ass}[basic assumptions]\label{ass:lipschitz}\label{ass:basic}
    \begin{enumerate}
        \item \label{ass:lipschitz:1} The functions $f_\ell ~\forall \ell \in [m]\coloneqq \{1,2,...,m\}$, and consequently $f$, in \eqref{eq:opt} are $L^0$-Lipschitz and $L$-smooth, \ie
                \begin{align*}
                    |f_\ell(\bm{\theta}) - f_\ell(\bm{\theta}')| &\leq L^0 \nrm{\bm{\theta}-\bm{\theta}'}, \quad \forall\bm{\theta},\bm{\theta}' \in \R^n;\\
                    \nrm{\nabla f_\ell(\bm{\theta}) - \nabla f_\ell(\bm{\theta}')} &\leq L \nrm{\bm{\theta}-\bm{\theta}'}, \quad \forall\bm{\theta},\bm{\theta}' \in \R^n.
                \end{align*} 
        \item \label{ass:lipschitz:unbiased} \label{ass:lipschitz:2} The stochastic zeroth-order oracle is unbiased: 
        \[
            f_\ell(\bm{\theta}) = \E[\bm{\xi} \sim \mathcal{D}^{\xi}_\ell]{F_\ell(\bm{\theta},\bm{\xi})}, ~\forall \ell \in [m], \forall \bm{\theta} \in \R^n.
        \]  
    \end{enumerate}
\end{ass}
\noindent
It is worth noting that, for the sake of simplicity, a common Lipschitz constant $L$ is assumed for all functions $f_\ell$.

As an example, the optimization problem in \eqref{eq:opt} deals with a scenario where there are $m$ agents, possibly with different tasks $\bm\theta_i \neq \bm\theta_j, i\neq j$. However, they also share a common task $f(\bm{\theta})$. 
Therefore, each agent $\ell$ seeks to optimize its local parameters $\bm\theta_\ell$ over the sum of all local cost functions.
This is different from multi-task or game scenarios where each agent $\ell$ addresses the problem given as $\minimize_{\bm\theta_\ell} f_\ell(\bm{\theta})$, with $f_\ell(\bm{\theta}) = g_\ell(\bm\theta_\ell) + R(\bm{\theta})$ consisting of the local cost $g_\ell$ and a regularizer $R$ capturing the relation between tasks $\bm\theta_j$ in multi-task scenarios \cite{nassif2020multitask} or $f_\ell(\bm{\theta})$ representing the local cost influenced by other players in the game \cite{narang_learning_2022}. 
In addition, it is different from federated learning \cite{mcmahan_communication-efficient_2017,li_federated_2020,yu_parallel_2019} or learning over graphs \cite{vlaski_distributed_2021-1,nassif_learning_2020,sayed_adaptation_2014,wu_decentralized_2017,duchi_dual_2012,hajinezhad2019zone} where agents seek a consensus in $\minimize_{\bm{\theta}=(\bm\theta_1,\cdots,\bm\theta_m)\in\R^n} \sum_{\ell=1}^{m} g_\ell(\bm\theta_\ell)$ by including the constraint $\bm\theta_\ell = \bm\theta_j, \ell \neq j$.
As an illustrative use-case, we focus on the relevant problem of RA in wireless communication systems.

\bigskip
\subsubsection*{\textbf{Application in RA}}
Consider the problem of RA where agents are represented as $m$ transmitters in the communication network, each serving a single receiver, with a total number of receivers $n\leq m$. The set of transmitters and receivers are denoted as $\mathcal{T}$ and $\mathcal{R}$, respectively. In the case where $m=n$, the communication network is considered ad-hoc, \ie each transmitter serves only one dedicated receiver. In the case where $m > n$, the receivers can be considered as base stations, where multiple transmitters have uplink communication with one base station, possibly causing interference to other base stations. We refer to $r(i) \in \mathcal{R}$ as the receiver dedicated to transmitter $i\in\mathcal{T}$. 
The communication channel between the $i$th transmitter and the $j$th receiver is considered as a random variable, denoted by $h_{ji} \in \mathbb{C}$. The matrix $\bm{H} \in \mathbb{C}^{|\mathcal{R}| \times |\mathcal{T}|}$, with $|\cdot|$ representing the cardinality of the set, contains all the channels in the communication network, with the channel $h_{ji}$ at the $j$th row and the $i$th column of the matrix.  

The RA problem aims to find optimal resources (\eg transmit power) for each user (as an agent) such that the global reward (\eg sum of user data rates for communication) is maximized. This is done by performing the optimization \eqref{eq:opt} with $f_\ell(\bm{\theta}) = -\E[\bm{\xi}=\bm{H}]{r^\pi_\ell(\bm\pi(\bm{\theta},\bm{H}),\bm{H})}$ where $\func{r^\pi_\ell}{\R^m \times \mathbb{C}^{|\mathcal{R}| \times |\mathcal{T}|}}{\R}$  denotes agent $\ell$'s reward or so-called utility, \eg, its data rate.
The vector $\bm\pi(\bm{\theta},\bm{H}) \coloneqq (\pi_1(\bm\theta_1,\bm{H}),\cdots,\pi_m(\bm\theta_m,\bm{H})) \in \R^m$ contains individual RA policies $\func{\pi_i}{\R^{n_i} \times \mathbb{C}^{|\mathcal{R}| \times |\mathcal{T}|}}{\R}$, \eg DNNs. 
These policies map their input to communication resources, \eg the transmit power of the corresponding agent. 
Each policy $\pi_i$ is parametrized by the policy parameter $\bm\theta_i \in \R^{n_i}$. 
Due to interference between agents through cross-talk channels $h_{ji},~j\neq i$, the utility of each agent, \ie $r^\pi_\ell$, depends on all the policies $\bm\pi$. Therefore, cooperative learning is required. Otherwise, the utilities are decoupled as $r^\pi_\ell(\pi_\ell(\bm\theta_\ell,h_{r(\ell) \ell}),h_{r(\ell) \ell})$ and the RA problem can be addressed trivially or by non-cooperative learning.

    \section{Proposed Learning Setup}\label{sec:setup}
        
In this section, we present the setup that governs the local updates and communication among the agents.
The frequently used notations are listed in \cref{tab:notations} in the Appendix.

\subsection{Agent Activity Time}\label{sec:setup:time}
Let $t\in\N$ represent the global time. 
The set of time instants at which agent $i$ updates its local parameters is defined as $T^i_u$.
Furthermore, each agent $i$ queries its zeroth-order oracle (\ie measures its cost function value) $F_i(\bm{\theta}^t,\bm{\xi})$ at certain time instants, which define the set $T^i_q$. 
At any time, each agent is free to decide whether to update its parameter or keep the current parameter, as well as whether to query its oracle or not. Consequently, we take the following assumption:
\begin{ass}[update and query times]\label{ass:timing}
    Updates and queries take place at certain time instants such that $T_q^i \subseteq \N$, $T_u^i \subseteq \N$ for all $i\in[m]$.
\end{ass}

\subsection{Message-passing Architecture}\label{sec:setup:comm}
\begin{figure}[t]
    \centering
    \begin{subfigure}{0.9\linewidth}
        \includegraphics[width=1\linewidth]{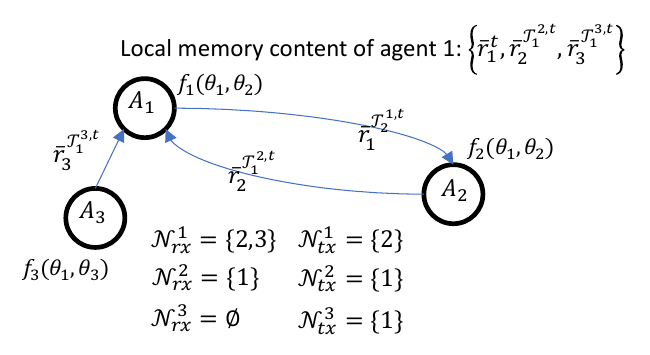}
        \caption{Message-passing architecture among the agents.
            The quantities $\bar r_j^t$ and $\mathcal{T}_{i}^{j,t}$ are defined in \eqref{eq:barr} and \eqref{eq:timestamp}, respectively.
            Depending on its local cost function, each agent defines the sets of agents $\mathcal{N}^i_{rx}$ and $\mathcal{N}^i_{tx}$ with which it communicates and saves necessary variables in its local memory. Due to asynchrony and random communication delays, the memory contents may be outdated.}
            \label{fig:messagePassing:a}
        \end{subfigure}
    \begin{subfigure}{0.9\linewidth}
        \includegraphics[width=1\linewidth]{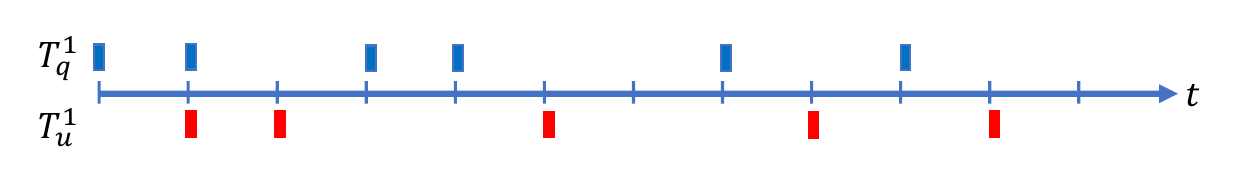}
        \caption{Update and query time of agent $1$.
            The axis ticks correspond to global time instants $t\in\N$. 
            Agents asynchronously query their cost functions and update their local parameters.}
            \label{fig:messagePassing:b}
    \end{subfigure}
    \caption{An example of a message-passing architecture with asynchronous agents. 
    The communication delays are possibly random and vary over time.
    The detailed workflow is provided in \cref{alg:proposed}.}
    \label{fig:messagePassing}
\end{figure}

The exchange of information among agents depends on the structure of the local cost functions $f_i$ for all $i\in[m]$.
To facilitate the presentation of this exchange, we introduce neighbor sets for each agent $i\in[m]$ as follows:
\begin{align}
    \begin{split}
        \mathcal{N}^i_{rx} &\coloneqq \{j\in [m] \mid f_j ~\text{depends on}~ \bm\theta_i\},\\
        \mathcal{N}^i_{tx} &\coloneqq \{j\in [m] \mid f_i ~\text{depends on}~ \bm\theta_j\}.
    \end{split}
    \label{eq:neighborset}
\end{align}
The significance of these neighbor sets will become clear in the subsequent sections.
The neighbor sets in \eqref{eq:neighborset} are application-dependent and agents are aware of them based on system design.
An example of a message-passing architecture with asynchronous agent activities is illustrated in \cref{fig:messagePassing}.

    \section{Zeroth-order optimization}
        
We utilize \emph{randomized smoothing techniques} \cite{duchi_2012_randomized,nesterov_random_2017} to leverage zeroth-order optimization methods in the proposed learning approach. %flaxman_2004_online
These techniques involve introducing random perturbations to smooth out the objective function.
Specifically, for a given function $f$ defined in \eqref{eq:opt}, its smooth \emph{approximation} at $\bm{\theta}$ is defined as
\begin{equation}
    f^\mu(\bm{\theta}) \coloneqq \E[\bm u \sim \mathcal{D}_u]{f(\bm{\theta} + \mu \bm u)},
    \label{eq:approximation_fun}
\end{equation}
where $\mu > 0$, and $\bm u$ is a random perturbation vector with a distribution $\mathcal{D}_u$.
Nesterov \cite{nesterov_random_2017} demonstrated that if the distribution is Gaussian, i.e., $\mathcal{D}_u = \mathcal{N}(\bm 0,\bm I)$, then $f^\mu$ exhibits the following properties:

\begin{prop}[the smooth approximation function $f^\mu$ \cite{nesterov_random_2017}]\label{prop:approximation}
    With the constant $\mu > 0$:
    \begin{enumerate}
        \item If $f$ is $L^0$-Lipschitz and $L$-smooth, $f^\mu$ is also $L^0_\mu$-Lipschitz and $L_{\mu}$-smooth with constants $L^0_\mu \leq L^0$ and $L_\mu \leq L$, \label{prop:approximation:1}
        \item The gradient $\nabla f^\mu$ is derived as
        \begin{equation*}
            \nabla f^\mu(\bm{\theta}) = \E[\bm u \sim\mathcal{N}(\bm 0,\bm I)]{\frac{f(\bm{\theta} + \mu \bm u) - f(\bm{\theta})}{\mu} \bm u},
        \end{equation*} \label{prop:approximation:2}
        \item For any $\bm{\theta} \in \R^n$,
            \begin{talign*}
                    |f^\mu(\bm{\theta}) - f(\bm{\theta})| &\leq \frac{\mu^2}{2} L n,\\
                    \|\nabla f^\mu(\bm{\theta}) - \nabla f(\bm{\theta})\| &\leq \frac{\mu}{2} L (n+3)^{\frac{3}{2}},\\
                    \E[\bm u]{\sqnrm{\frac{f(\bm{\theta} + \mu \bm u) - f(\bm{\theta})}{\mu} \bm u}} &\leq 2(n+4)\sqnrm{\nabla f(\bm{\theta})} \\
                    &+ \frac{\mu^2}{2}L^2(n+6)^3.
            \end{talign*} \label{prop:approximation:3}
    \end{enumerate}
\end{prop}
The technique is called randomized smoothing, as the approximation function $f^\mu$ is smooth (refer to \cref{ass:basic} for definition) even if the function $f$ is not \cite[lem. 2]{nesterov_random_2017}.
The computation of the gradient $\nabla f^\mu$ in \cref{prop:approximation:2} constitutes the key computational task for each agent in the proposed learning approach, as elaborated in the following section. 
Importantly, this gradient derivation solely relies on querying the zeroth-order oracle, \ie the cost function value, using perturbed and unperturbed decision variables. 
It is noteworthy that throughout the learning process, the smooth approximation function $f^\mu$ is optimized, rather than the original function $f$. 
However, as indicated by \cref{prop:approximation:3}, it becomes evident that by selecting a sufficiently small constant $\mu$, the two functions can be made arbitrarily close to each other.

    \section{Proposed Distributed learning}\label{sec:proposed_alg}
        In this section, we propose a distributed learning approach within a message-passing architecture. 
In this framework, agents collaborate to optimize their local policy parameters $\bm\theta_i$ gathered in $\bm{\theta} = (\bm\theta_1, \cdots, \bm\theta_m)$ in \eqref{eq:opt}.

In accordance with \cref{prop:approximation:2}, each agent $i$ can perform queries at time $t\in T_q^i$ by measuring its local cost function values and using the following expression:
\begin{align*}
    \bar{r}_i^t \coloneqq \frac{1}{\mu m B} &\sum_{j=1}^B  \left\{F_i(\bm{\theta}^t,\bm{\xi}^{t,j}) - F_i(\bm{\theta}^t+\mu \bm u^t,\bar{\bm{\xi}}^{t,j})\right\} \in \R,\\
    \text{where}~~ &\bm u^t=(\bm u_1^t,\cdots,\bm u_m^t) \sim \mathcal{D}_u=\mathcal{N}(\bm 0, \bm I_{n \times n}). \label{eq:barr} \numberthis
\end{align*}
Here, $t\in\N$ represents the global time, $F_i$ is the cost function value (oracle) defined in \cref{ass:lipschitz:2}, and the summations are over two minibatches of sample vectors $\bm{\xi}^{t,j}$ and $\bar{\bm{\xi}}^{{t,j}}$ for $j\in[B]$.  
The queries in \eqref{eq:barr} with perturbed and unperturbed parameters can be performed in two different time slots, possibly using distinct sample vectors.
While the oracle $F_i$ of agent $i$ relies on the (perturbed) parameters of its neighbors $j\in\mathcal{N}_{tx}^i$, agent $i$ does not require receiving the parameters $\bm\theta_j$. It only needs to query its own oracle $F_i$, affected by $\bm\theta_j$ through their interaction.

It is essential to highlight that the query in \eqref{eq:barr} requires a certain level of synchrony among the agents. 
Specifically, the oracles $F_i(\bm{\theta}^t,\bm{\xi}^{t,j})$ and $F_i(\bm{\theta}^t+\mu \bm u^t,\bar{\bm{\xi}}^{t,j})$ necessitate the alignment of parameters $\bm\theta_i$ and $\bm\theta_i+ \mu \bm u_i$ in time for all agents $i\in[m]$ to ensure the validity of the queries. 
This implies that the agents should synchronously employ their perturbed and unperturbed local parameters, even if $t \notin T_q^i$. This requirement arises solely from employing zeroth-order oracles and can be lifted if higher-order oracles (\eg function gradients) are utilized. 
Nonetheless, by \cref{ass:timing}, agents still have the flexibility to asynchronously choose their preferred times $T_u^i, T_q^i$ for updating, querying their oracles, and communicating with other agents, based on their heterogeneous communication and computational capabilities.

To update its parameter, each agent $i$ requires not only its own queries $\bar{r}_i^t$ but also queries from agents whose performance depends on $\bm\theta_i$, namely the agents in the set $\mathcal{N}^{i}_{rx}$ defined in \eqref{eq:neighborset}. 
Parameter sharing is achieved through message-passing, which is susceptible to random (but bounded) communication delays. 
Therefore, each agent $i$ updates its parameter using potentially outdated queries received from its neighbors $\mathcal{N}^{i}_{rx}$.
To establish the age of the queries, with a maximum $D^{\max} \geq 0$, consistent with the seminal work in \cite{bertsekas_parallel_2015}, we introduce the timestamp $t-D^{\max} \leq \mathcal{T}_{i}^{\ell,t} \leq t$ defined as follows:
\begin{align}
    \begin{split}
    &\text{$\mathcal{T}_{i}^{\ell,t}$: the global time at which agent $\ell$ has queried}\\
    &\text{its~ oracle and ~the corresponding ~query is stored}\\
    &\text{in agent $i$'s local memory at time $t$.}
    \end{split}
    \label{eq:timestamp}
\end{align}
It is evident that $\mathcal{T}_{i}^{i,t}=t, \forall i\in[m]$, $\mathcal{T}_{i}^{\ell,t}\in T_q^\ell$, and $0 \leq t - \mathcal{T}_{i}^{\ell,t} \leq D^{\max}$ indicates how old agent $\ell$'s query stored in agent $i$'s memory is, with a maximum age of $D^{\max}$. An example of a message-passing architecture and the involved quantities are depicted in \cref{fig:messagePassing}.

Utilizing the quantities $\bar{r}_\ell^{\mathcal{T}_{i}^{\ell,t}}$ for $\ell\in\mathcal{N}^{i}_{rx}$ along with the vectors $\bm u_i^{\tau}$ for $\tau \in \{t-D^{\max},\cdots,t\}$, stored in the local memories $\mathcal{B}_i$ and $\tilde{\mathcal{B}_i}$ respectively, each agent $i$ performs an asynchronous update at the time of interest $t\in T_u^i$ following \cref{alg:proposed}. 
These updates are executed using stepsize sequences $\seq{\gamma^t_i}[{t\in\N}]$.
To properly align the queries $\bar{r}_\ell^{\mathcal{T}_{i}^{\ell,t}}$ with the random perturbations $\bm u_i^{\mathcal{T}_{i}^{\ell,t}}$ in time, as required in \eqref{eq:barr}, each agent $i$ needs to know the global time $\mathcal{T}_{i}^{\ell,t}$ associated with the queries $\bar{r}_\ell^{\mathcal{T}_{i}^{\ell,t}}$ received from its neighbors $\ell\in\mathcal{N}^{i}_{rx}$. 
Therefore through message-passing, this global time is also exchanged, along with the corresponding query, to ensure the required alignment.

\begin{algorithm}[t]
    \begin{algorithmic}[1]
    \item[\algfont Initialize:] 
        \begin{tabular}[t]{@{}l@{}} 
            $(\gamma_i^t)_{t=0}^{\bar t-1},~ \bm\theta_i^0 \in \R^{n_i}, ~\bar{r}_\ell^0 = 0, \mathcal{T}_{i}^{\ell,0}=0, \ell\in\mathcal{N}^{i}_{rx}$,\\
            $\mathcal{D}_{u_i} = \mathcal{N}(\bm 0,\bm I), \bm u_i^0 \sim \mathcal{D}_{u_i}, \mu > 0, B > 0$,\\
            $T_q^i, T_u^i, ~i \in [m]$
        \end{tabular}

    \vspace{6pt} 
    \item[\algfont For $t=0,\cdots,\bar t-1$, each agent $i$ asynchronously:] 
    \item[\textit{$\blacktriangleright$ Quering and broadcasting:}]
    \vspace{6pt} 
    \State If $t\in T_q^i$: 
    \begin{algsubstates}
        \item compute $\bar{r}_i^t$ using \eqref{eq:barr} with $\bm\theta_i^t$ and $\bm u_i^t$
        \item update the memory $\mathcal{B}_i$ by $\{\bar{r}_i^t, t\}$ 
        \item send $\{\bar{r}_i^t, t\}$ to the neighbors $j\in \mathcal{N}^i_{tx}$ \label{alg:sending}
    \end{algsubstates}
    \item[\textit{$\blacktriangleright$ Local memory updates:}]
    \vspace{6pt}  
    \State \label{alg:receiving} Receive new updates from the neighbors $\ell\in\mathcal{N}^{i}_{rx}$ denoted by $Q_\ell=\{r_\ell,\mathcal{T}_\ell\}$
    \State For each $\ell\in\mathcal{N}^{i}_{rx}$, if $Q_\ell \neq \emptyset$: 
    \begin{algsubstates}
        \item update the memory $\mathcal{B}_i$ by $\mathcal{T}_{i}^{\ell,t} \gets \mathcal{T}_\ell, ~ \bar{r}_\ell^{\mathcal{T}_{i}^{\ell,t}} \gets r_\ell$
    \end{algsubstates}
    \State $\mathcal{T}_{i}^{\ell,t+1} = \mathcal{T}_{i}^{\ell,t}, \quad \forall \ell\in\mathcal{N}^{i}_{rx}$
    \State If the memory $\tilde{\mathcal{B}_i}$ is full, omit the oldest entry
    \State $\bm u_i^{t+1} \sim \mathcal{D}_{u_i}$ and save it in the local memory $\tilde{\mathcal{B}_i}$
    \item[\textit{$\blacktriangleright$ Local updates:}]
    \vspace{6pt}  
    \State \label{alg:updating} Using the locally stored $\bm u_i^{\tau}$ for $\tau \in \{t-D^{\max},\cdots,t\}$ and $\bar{r}_\ell^{\mathcal{T}_{i}^{\ell,t}}, ~\ell\in\mathcal{N}^i_{rx}$ perform
        \begin{align}
            \begin{split}
                \bm\theta_i^{t+1} = \bm\theta_i^{t} + \gamma^{t}_i \bm{\tilde{s}}_i^{t}
                % \begin{cases}
                %     \bm\theta_i^{t-1} + \gamma^{t-1}_i \bm{\tilde{s}}_i^{t-1}, &~~\text{if}~~~ t\in T_u^i,\\
                %     \bm\theta_i^{t-1}, &~~\text{if}~~~ t\notin T_u^i,
                % \end{cases}
            \end{split}
            \label{eq:async:individual}
        \end{align}
        where $\bm{\tilde{s}}_i^t$ is defined by
        \begin{align}
                \bm{\tilde{s}}_i^t \coloneqq %\textstyle\sum_{\ell\in\mathcal{N}^i_{rx}} \bar{r}_\ell^{\mathcal{T}_{i}^{\ell,t}} \bm u_i^{\mathcal{T}_{i}^{\ell,t}},
                \begin{cases}
                    \textstyle\sum_{\ell\in\mathcal{N}^i_{rx}} \bar{r}_\ell^{\mathcal{T}_{i}^{\ell,t}} \bm u_i^{\mathcal{T}_{i}^{\ell,t}}, &~~\text{if}~~~ t\in T_u^i,\\
                    \bm 0, &~~\text{if}~~~ t\notin T_u^i,
                \end{cases}
            \label{eq:s}
        \end{align}
        with $\mathcal{T}_{i}^{\ell,t}$ in \eqref{eq:timestamp}
\end{algorithmic}

    \caption{Proposed asynchronous distributed learning}
    \label{alg:proposed}
\end{algorithm}

In \cref{alg:proposed}, it is evident that utilizing zeroth-order oracles by each agent has three advantages. 
First, the proposed distributed learning approach can handle cost functions for which higher-order oracles (\eg function gradients) may not be available or are expensive to calculate (\eg in simulation-based optimizations or bandit optimizations \cite{hajinezhad2019zone}). Second, unlike federated learning or learning over graph settings where agents share local parameter or gradient vectors, employing zeroth-order oracles significantly reduces the communication bandwidth requirements in \cref{alg:proposed} as agents only share scalars, by which the required gradients can be estimated (refer to \cref{alg:sending} and \cref{alg:receiving} for message-passing; \cref{prop:approximation:2,alg:updating} for gradient estimation). Third, sharing scalars adds a layer of privacy to distributed learning, mitigating the risk of privacy leakage, a known issue in federated learning settings. \cite{wang2019beyond,hitaj2017deep}.

An essential requirement for ensuring the convergence of \cref{alg:proposed} is to establish an upper bound for $D^{\max}$, \ie the maximum age of queries in each agent's local memory. This necessitates imposing appropriate assumptions on the message-passing architecture and the activity time of agents that define $D^{\max}$.
We present these crucial assumptions in the subsequent section, along with additional necessary assumptions on the zeroth-order oracles.
    \section{Convergence study}
        \subsection{Further definitions and assumptions}\label{sec:defs}
            To facilitate the convergence study, we define the following vectors associated with agent $i$ at time $t$:
\begin{align}
    \begin{split}
        \Theta^t_i &\coloneqq \vect\{\bm{\theta}^{\mathcal{T}_{i}^{\ell,t}} \in \R^n \mid \ell\in \mathcal{N}^{i}_{rx}\} \in \R^{n|\mathcal{N}^{i}_{rx}|}, \\
        \mathbb{U}^t_i &\coloneqq \vect\{\bm u^{\mathcal{T}_{i}^{\ell,t}} \in \R^n \mid \ell \in \mathcal{N}^{i}_{rx}\} \in \R^{n|\mathcal{N}^{i}_{rx}|},
    \end{split}
    \label{eq:Params}
\end{align}
where $\vect\{\cdot\}$ denotes the vectorization operator. 
Additionally, we define the following functions using the variables defined in \eqref{eq:Params}:
\begin{talign*}
    \tilde{f}_i(\Theta_i^t) &\coloneqq \frac{1}{m} \sum_{\ell\in\mathcal{N}^{i}_{rx}} f_\ell(\bm{\theta}^{\mathcal{T}_{i}^{\ell,t}}), \\
    \tilde{f}^\mu_i(\Theta_i^t) &\coloneqq \frac{1}{m} \sum_{\ell\in\mathcal{N}^{i}_{rx}} f^\mu_{\ell}(\bm{\theta}^{\mathcal{T}_{i}^{\ell,t}}), \label{eq:s_expect}\numberthis\\
    f^\mu_{\ell}(\bm{\theta}^{\mathcal{T}_{i}^{\ell,t}}) &\coloneqq \E[\bm{\xi},\bm u]{F_\ell(\bm{\theta}^{\mathcal{T}_{i}^{\ell,t}} + \bm u,\bm{\xi})}.
\end{talign*}
The functions defined in \eqref{eq:s_expect} represent the objective functions by which agent $i$ optimizes its local parameter $\bm\theta_i$. Based on \eqref{eq:barr}, \eqref{eq:s}, \cref{ass:lipschitz:2}, and \cref{prop:approximation:2}, it can be observed (refer to further discussion in \cref{lem:unbiased} in Appendix) that the update $\bm{\tilde s}_i^t$ in \cref{alg:proposed} satisfies
\begin{equation}
     \E[\bm{\xi},\mathbb{U}]{\bm{\tilde s}_i^t} = -\nabla_i \tilde{f}^\mu_i(\Theta_i^t), \quad t\in T_u^i,
    \label{eq:s_new_mean}
\end{equation}
where $\nabla_i$ represents the gradient with respect to $\bm\theta_i \in \R^{n_i}$ defined as
\begin{equation}
    \textstyle\nabla_i \tilde{f}^\mu_i(\Theta_i^t) \coloneqq \frac{1}{m} \sum_{\ell\in\mathcal{N}^{i}_{rx}} \nabla_if^\mu_{\ell}(\bm{\theta}^{\mathcal{T}_{i}^{\ell,t}}).
    \label{eq:gradDef}
\end{equation}
By considering \eqref{eq:s_new_mean}, each agent $i$ updates its local parameter $\bm\theta_i$ using the function $\tilde{f}^\mu_i(\Theta_i^t)$ with potentially outdated parameters $\Theta_i^t$. 
In the subsequent section, we demonstrate that these outdated parameters eventually converge to a common parameter $\bm{\theta}^t$ with possibly different local parameters.
To show this convergence, let us introduce another useful parameter associated with agent $i$:
\begin{equation}
    \bar\Theta^t_i \coloneqq (\overbrace{\bm{\theta}^t,\cdots,\bm{\theta}^t}^{|\mathcal{N}^{i}_{rx}| ~\text{times}}) \in \R^{n|\mathcal{N}^{i}_{rx}|}.
    \label{eq:theta_common}
\end{equation}
In an ideal convergence scenario, we seek for $\Theta_i^t \to \bar\Theta^t_i$ (refer to \cref{thm:funval}), where the following holds for any $i\in[m]$:
\begin{align}
    \begin{split}
        \nabla_i \tilde{f}_i(\bar\Theta^t_i) &= \nabla_i f(\bm{\theta}^t),\\
        \nabla_i \tilde{f}^\mu_i(\bar\Theta^t_i) &= \nabla_i f^\mu(\bm{\theta}^t), % = -\E[\bm{\xi},\bm u]{\bm s_i^t},
    \end{split}
    \label{eq:funcs:theta_common}
\end{align}
with $f^\mu$ defined in \eqref{eq:approximation_fun}. % and $\bm s_i^t$ is defined in \eqref{eq:barr}.
Moreover, leveraging \eqref{eq:async:individual} and \eqref{eq:s}, it is useful to formulate the joint update of local parameters as:
\begin{equation}
    \bm{\theta}^{t+1} = \bm{\theta}^t + \Gamma^t \bm {\tilde s}^t, \quad t\in\N,
    \label{eq:update_rule}
\end{equation}
where $\bm {\tilde s}^t \coloneqq (\bm{\tilde s}_1^t,\cdots,\bm{\tilde s}_m^t) \in \R^n$ and $\Gamma^t \coloneqq \diag\{\gamma^t_1 \I_{n_1\times n_1},\cdots,\gamma^t_m \I_{n_m\times n_m}\}$ with identity matrices $\I$ with appropriate dimensions.

Ensuring a bounded delay is a crucial assumption in our convergence analysis (refer to \cref{thm:subseq}).
To establish such a guarantee, we impose the following assumption on the sets $T_u^\ell$ and $T_q^\ell$ in consistent with \cite[Sec. 7]{bertsekas_parallel_2015}:
\begin{ass}[asynchrony level]\label{ass:asynchrony}
    There exists a finite $D > 0$ such that for all $\ell\in[m]$ and $t\in\N$, at least one of the elements in the set $\{t,t+1,\cdots,t+D-1\}$ belongs to the sets $T^\ell_q$ and $T^\ell_u$. Moreover, the communication delay is upper bounded by $D$. \label{ass:asynchrony:1}
\end{ass}
\noindent
\cref{ass:asynchrony} ensures that all agents actively participate in the learning process and that no agent remains idle indefinitely before the learning is completed.
Based on this assumption, we can derive the following upper bound for the maximum age of received quantities $\bar{r}_\ell$:
\begin{prop}[maximum delay]\label{prop:globalInfoMaxDel}
    There exists a finite $D^{\max} \geq D$ such that for all $i\in[m]$, $\ell\in\mathcal{N}^{i}_{rx}$, and $t\in\N$, there holds $t-D^{\max} \leq \mathcal{T}_{i}^{\ell,t} \leq t$ \label{ass:asynchrony:2}.
\end{prop}
\noindent
\cref{prop:globalInfoMaxDel} states that agents update their parameters $\bm\theta_i^t \in \R^{n_i}$ using the outdated quantities $\bar{r}_\ell^{\mathcal{T}_{i}^{\ell,t}}, \ell\in\mathcal{N}^{i}_{rx}$ defined in \eqref{eq:barr}, which are at most $D^{\max}$ time instants old.
With $D^{\max}$ in \cref{prop:globalInfoMaxDel}, we introduce the next necessary assumption:
\begin{ass}[local memory size]\label{ass:buffer:size}
    Each agent $i \in [m]$ has the memory $\tilde{\mathcal{B}_i}$ of size $D^{\max}+1$ to save the local perturbation vectors $\bm u_i^\tau \in \R^{n_i}$ for $t - D^{\max} \leq \tau \leq t$.
\end{ass}
\noindent
\cref{ass:buffer:size} ensures that agents can compute $\bm{\tilde{s}}_i^t$ in \eqref{eq:s} using the saved perturbation vectors. 
It is remarked that the perturbation vectors are generated and saved locally, and they are not shared through message-passing.

Finally, we present the following required assumptions on the function gradients, oracles, and sample vectors:
\begin{ass}[boundedness]\label{ass:oracle}
    \begin{enumerate}
        \item $\E[\bm{\xi}]{\sqnrm{\nabla F_\ell(\bm{\theta}, \bm{\xi}) - \nabla f_\ell(\bm{\theta})}} \leq \sigma^2, \forall \ell\in[m], \forall \bm{\theta} \in \R^n$, \label{ass:oracle:1}
        \item $\nrm{\nabla f_\ell(\bm{\theta})} \leq G, ~\forall \ell\in[m], \forall \bm{\theta} \in \R^n$, \label{ass:oracle:2}
    \end{enumerate}
\end{ass}
\begin{ass}[sample vector]\label{ass:oracle:noise}
    \begin{enumerate}
        \item\label{ass:oracle:3} $\E[\bm{\xi}]{\sqnrm{\bm{\xi}}} \leq v^2$; 
        \item\label{ass:lipschitz:3}  The stochastic zeroth-order oracle is Lipschitz continuous in $\bm{\xi}$, $\forall \ell \in [m]$ and $\forall\bm{\theta} \in \R^n$:
        \[
            \nrm{F_\ell(\bm{\theta},\bm{\xi}) - F_\ell(\bm{\theta},\bm{\xi}')} \leq L_{\bm{\xi}} \nrm{\bm{\xi} - \bm{\xi}'} ,~ \forall \bm{\xi}, \bm{\xi}' \in \R^s.
        \]
    \end{enumerate}
\end{ass}
\noindent
\cref{ass:oracle:1,ass:oracle:2} are standard in the stochastic optimization literature \cite{ghadimi_mini-batch_2016}. 
Additional \cref{ass:oracle:3,ass:lipschitz:3}, however, are required specifically concerning the sample vector.
The additional assumptions are necessary because the quantity $\bar{r}_i^t$ in \eqref{eq:barr} is obtained using different sample vectors for the perturbed and unperturbed functions. 
While most existing works in literature, \eg \cite{ghadimi2013stochastic,ghadimi_mini-batch_2016,hajinezhad2019zone,kalogerias_model-free_2020}, assume that the oracle $F_\ell(\cdot, \bm{\xi})$ can be queried at two different points $\bm{\theta}$ and $\bm{\theta}'$ in $\R^n$ with a single sample vector $\bm{\xi}$, this assumption does not hold in many practical applications, including the use-case discussed in \cref{sec:sim}. In these applications, allowing different sample vectors $\bm{\xi}$ at different points $\bm{\theta}$ makes the implementation of zeroth-order optimization possible, as queries can be taken at separate time slots.
We will further examine the impact of this practical limitation on the convergence of \cref{alg:proposed} in the subsequent section (refer to \cref{cor:twoPointQuery}), building upon studies of asynchronous parallel and distributed computations presented in \cite{bertsekas_parallel_2015}.
        \subsection{Main results}\label{sec:main_results}
            The subsequent results provide proof of convergence and convergence rate for the proposed asynchronous distributed learning approach in \cref{alg:proposed}, addressing problem \eqref{eq:opt}.
In the first theorem, we show that a subsequence of $\seq{\Theta^t_\ell}[t\in\N]$, \ie of the outdated local parameters, converges to a neighborhood of some stationary point of the objective function $\tilde{f}_\ell$ by which agent $\ell$ optimizes its local parameter.
\begin{thm}[subsequential convergence]\label{thm:subseq}
    Take \cref{ass:timing,ass:lipschitz,ass:asynchrony,ass:buffer:size,ass:oracle,ass:oracle:noise}, the smoothing parameter $\mu>0$ defined in \cref{prop:approximation}, the batch size $B$ for all users in \eqref{eq:barr}, and the stepsize sequences $\seq{\gamma^t_i}[t\in\N] \in (0,\nicefrac{1}{M}), i\in[m]$ where
    \begin{align*}
         M \coloneqq \nicefrac{L(2m D^{\max} + 1)}{2}.
    \end{align*}
    Then for the iterates generated by \cref{alg:proposed} at $\{0,1,\cdots,\bar t\}$, the following holds for all $\ell\in[m]$:
    \begin{align*}
        &\min_{t \in \{0,\cdots,\bar t\}} \E{\sqnrm{\nabla_\ell\tilde{f}_\ell(\Theta^{t}_\ell)}} \leq \frac{1}{2m} \mu^2 L^2(n+3)^{3} \label{eq:convRate} \numberthis\\
        &+ \frac{2\mu^2Ln + 2(f(\bm{\theta}^0)-f^\star) + 2M \frac{\tilde\sigma^2}{B} \sum_{t=0}^{\bar t}\sum_{\ell=1}^m (\gamma^t_\ell)^2}{\eta \sum_{t=0}^{\bar t} \bar\gamma^t},
    \end{align*}
    where 
    \begin{align*}
        \eta &\coloneqq 1 - M \gamma^{\max}, ~
        \gamma^{\max} \coloneqq \max_{i\in[m],t\in\N}[\gamma^t_i], ~
        \bar\gamma^t \coloneqq \min_{i\in[m]}[\gamma^t_i] \forall t,\\
        \tilde\sigma^2 &\coloneqq 4(n+4) (\sigma^2 + G^2) + {\mu^2}L^2(n+6)^3 + 4\frac{B L^2_{\bm{\xi}}}{\mu^2} v^2,
    \end{align*}
    and $f^\star$ is the minimum of the function $f$.
    \begin{proof}
        Refer to Appendix \ref{sec:app:conv}.
    \end{proof}
\end{thm}
The same dependency on the problem dimension $n$ is observed in existing works on nonconvex zeroth-order settings, \eg \cite{hajinezhad2019zone,nesterov_random_2017,ghadimi_mini-batch_2016,ghadimi2013stochastic}. Furthermore, the stepsize range provided in \cref{thm:subseq} is twice as large as the range reported in \cite{ghadimi_mini-batch_2016}, when there is no delay, \ie $D^{\max} = 0$.
However, it is evident from the defined parameter $M$ that the introduction of delay due to asynchrony and communication delays requires the stepsize to decrease linearly with both the number of agents $m$ and the delay in order to ensure convergence.

An important distinction in the complexity \eqref{eq:convRate} lies in the presence of the last term in the variance term $\tilde\sigma^2$ which makes the tuning of variable $\mu$ challenging. In \eqref{eq:convRate}, it is evident that a smaller $\mu$ results in a smaller first term in the right-hand side; however, it also increases the variance $\tilde\sigma^2$. In \cref{thm:funval}, we will see that this results in convergence to a point closer to the stationary point but at a slower rate. It is essential to note that this term arises solely from the practical constraint that prohibits querying the oracle in two different points with the same sample vector. The following corollary establishes the iteration complexity when this constraint is lifted. The proof of this corollary is omitted here due to space limitations.
\begin{cor}\label{cor:twoPointQuery}
    Take \cref{ass:timing,ass:lipschitz,ass:asynchrony,ass:buffer:size,ass:oracle}. In addition, it is assumed that the oracles $F_\ell(\cdot,\bm{\xi}), \forall \ell\in[N]$ can be queried in any two points $\bm{\theta},\bm{\theta}' \in \R^n$. Hence, the queries in \eqref{eq:barr} are performed by similar sample vectors in perturbed and unperturbed oracles for each sample $j$. 
    Then for the iterates generated by \cref{alg:proposed} the iteration complexity in \eqref{eq:convRate} holds with the sample variance $\tilde\sigma^2 \coloneqq 8(n+4) (\sigma^2 + G^2) + 2{\mu^2}L^2(n+6)^3$.
\end{cor}
\noindent

We proceed to demonstrate the convergence of the original function $f$ in \eqref{eq:opt}, the outdated parameters $\bm{\theta}^{\mathcal{T}_{\ell}^{i,t}}$ to a common parameter $\bm{\theta}^t$, and the function values $f_i(\bm{\theta}^{\mathcal{T}_{\ell}^{i,t}})$.
\begin{thm}[asymptotic convergence and convergence rate]\label{thm:funval}
    Take the assumptions in \cref{cor:twoPointQuery} and that the steptizes are diminishing such that $\lim_{t\to\infty} \gamma_i^t = 0, i\in[m]$. Then for all $\ell\in[m]$: 
    \begin{enumerate}
        \item $\lim_{t\to\infty} \E{\sqnrm{\bar\Theta^t_\ell - \Theta^t_\ell}} = 0$; \label{thm:funval:1}
        \item $\lim_{t\to\infty} \E{|f_i(\bm{\theta}^t) - f_i(\bm{\theta}^{\mathcal{T}_{\ell}^{i,t}})|} = 0, ~~\forall i \in \mathcal{N}^{\ell}_{rx}$. \label{thm:funval:3}
    \end{enumerate}
    Moreover, if the diminishing stepsize sequences are defined by $\gamma^t_i = \nicefrac{\gamma^0_i}{\sqrt{t+r}}$ with some $r > 0$,
    then
    \begin{equation}
        \min_{t\in\{0,\cdots,\bar t\}}\E{\sqnrm{\nabla f(\bm{\theta}^t)}} \leq \mathcal{O}(\mu^2) + \mathcal{O}(\nicefrac{1}{\sqrt{\bar t}}).
        \label{thm:funval:2}
    \end{equation}
    \begin{proof}
        Refer to Appendix \ref{sec:app:conv}.
    \end{proof}
\end{thm}
\noindent
According to \eqref{thm:funval:2}, a subsequence of $\seq{\bm{\theta}^t}[t \in\N]$ converges to a neighborhood of some stationary point of $f$, and the size of this neighborhood is determined by the value of $\mu$. 
This convergence rate is consistent with existing works addressing zeroth-order nonconvex problems with synchronous updates, \eg \cite{ghadimi2013stochastic,ghadimi_mini-batch_2016,hajinezhad2019zone}. The rate $\mathcal{O}(\nicefrac{1}{\sqrt{\bar t}})$ can also be achieved if the parameter $\mu$ diminishes proportionally to $\mathcal{O}(\nicefrac{1}{\sqrt{\bar t}})$ \cite{ghadimi2013stochastic,ghadimi_mini-batch_2016}. In addition, if the batch size increases proportionally to $\mathcal{O}(\sqrt{t})$, the rate $\mathcal{O}(\nicefrac{1}{t})$ is achieved \cite{hajinezhad2019zone}.
If the assumption that the oracle $F_\ell(\cdot, \bm{\xi})$ can be queried at two different points $\bm{\theta}$ and $\bm{\theta}'$ in $\R^n$ with a single sample vector $\bm{\xi}$ does not hold, which is the case in many practical applications, including the use-case considered in \cref{sec:sim}, the convergence rate in \eqref{thm:funval:2} becomes $\mathcal{O}(\mu^2) + \mathcal{O}(\nicefrac{1}{(\mu^2\sqrt{\bar t})})$, which can be concluded from the proof in Appendix \ref{sec:app:conv}. Hence, in this case, a smaller $\mu$ guarantees convergence to a tighter neighborhood of some stationary point; however, the convergence rate becomes slower due to the last term.

    \section{Implementation of the proposed distributed learning approach in RA problem}\label{sec:RA}
        In this section, we explore the application of the proposed distributed learning in DL-based RA for communication networks, as an example (refer to \cref{sec:system_model} for a brief introduction). 
In the considered RA scenario, we aim to maximize the sum utility of agents (representing the global reward function) in the communication network, which can be formulated by \eqref{eq:opt}.
Since typical utility/loss functions in RA, such as data rate, communication latency, and appropriate variants that guarantee fairness, energy efficiency, etc., satisfy the assumptions considered in \cref{ass:basic,ass:oracle,ass:oracle:noise}, the learning process can be addressed by the proposed learning approach in a distributed manner, benefiting from the advantages mentioned in \cref{sec:intro:B}.
It is important to note that for the considered example, in practice, each agent $i$, acting as a transmitter, can have a significant impact on the utility of a large number of agents due to cross-talk channels. 
For the purposes of this section, we assume that each agent $i$ influences the utility of all the other agents.
Therefore, the neighbor sets defined in \eqref{eq:neighborset} are $\mathcal{N}^{i}_{rx} = [m]$ and $\mathcal{N}^i_{tx}=[m]$, necessitating a fully connected message-passing architecture where each agent communicates with all other agents, which is not always a practical solution. In this case, agent $i$ may choose to communicate with only a few agents (immediate neighbors) to gather the necessary information. 
Hence, the queries $\bar{r}_\ell, \forall \ell\in[m]$ required by each agent $i$ are passed through intermediate agents and immediate neighbors (possibly with delays) to reach agent $i$. 
Consequently, in this application, it is necessary to redefine the message-passing architecture among agents and to establish new sets of neighbors.
        \subsection{Setup}\label{sec:setup:RA}
            
The agent activity time is the same as the one defined in \cref{sec:setup:time}.
We now establish new sets of neighbors.
\newline
\subsubsection{Underlying graph}\label{sec:RAsetup:graph}
Let us consider a directed graph $\mathcal{G}(\mathcal{V},\mathcal{E})$, where the set of vertices is denoted as $\mathcal{V} \coloneqq [m]$ and the corresponding edges are represented by $\mathcal{E} \coloneqq \{a_{ij} \in \{0,1\} \mid i,j \in [m] \}$. 
In this graph, we set $a_{ij}=1$ if agent $i$ transmits the necessary quantities to agent $j$, and $a_{ij}=0$ if there is no communication from agent $i$ to agent $j$. It is important to note that $a_{ij}$ can be different from $a_{ji}$. The neighbor sets of agent $i$, consisting of the agents that send and receive quantities to and from agent $i$, respectively, are defined as follows:
\begin{align}
    \begin{split}
        \bar{\mathcal{N}}^i_{rx} &\coloneqq \{j\in[m] \mid a_{ji}=1\},\\
        \bar{\mathcal{N}}^i_{tx} &\coloneqq \{j\in[m] \mid a_{ij}=1\}.
    \end{split}
    \label{eq:RA_neighbor_sets}
\end{align}

\noindent
The underlying graph is assumed to be fixed and depends on practical considerations, \eg existing backhaul communication links, etc.
In the proposed learning approach, it is essential for the underlying graph to be connected.
\begin{ass}[connected graph $\mathcal{G}$ \cite{sayed_adaptation_2014}]\label{ass:graph}
    The underlying graph $\mathcal{G}$ is \emph{connected} such that any two vertices are linked in both directions either directly or through other vertices. 
    The two directions possibly are two different paths in the graph. 
\end{ass}

\subsubsection{Communication between the agents}\label{sec:RAsetup:comm}
In the considered application, each agent affects the performance of all other agents through interaction (via cross-talk channels in the communication network). Therefore, it is important that each agent gathers the quantities $\bar{r}_\ell$ from all agents. 
It is more practical if agent $i$ engages in communication with its neighbors $\bar{\mathcal{N}}^i_{rx}$ and $\bar{\mathcal{N}}^i_{tx}$ only at specific times of interest during the learning process. The time instants for communication are defined by the set $T^i_{tx}$ for agent $i$. 
At a given time $t\in T^i_{tx}$, agent $i$ transmits all the quantities $\bar{r}_\ell^{\mathcal{T}_{i}^{\ell,t}}, \ell\in[m]$ (global information) stored in its memory to its neighbors $j \in \bar{\mathcal{N}}^i_{tx}$. 
The communication delay within the graph is random and bounded.

Since the graph $\mathcal{G}$ is connected, two agents may be connected through multiple paths. This means that agent $i$ may receive multiple quantities from agent $\ell\neq i$ with possibly different delays. 
In such cases, agent $i$ has the option to discard the older quantities and retain the most recent ones.
Hence, it is crucial for each quantity to be associated with an agent ID as well. 
Therefore, throughout the learning process, each agent $i$ stores and sends its query $\bar{r}_i^t$ along with a timestamp $t$ and the agent ID $i$. Similarly, the received quantities $\bar{r}_\ell^{\mathcal{T}_{i}^{\ell,t}}$ from the neighboring agents $j\in\bar{\mathcal{N}}^i_{rx}$ are also stored in the memory $\mathcal{B}_i$ along with their corresponding timestamps $\mathcal{T}_{i}^{\ell,t}$ and agent IDs $\ell\in[m]$.
        \subsection{Algorithm}\label{sec:alg:RA}
            The learning process requires the active participation of all agents to ensure that no agent is isolated and that each agent can transmit and receive the necessary information. This cooperative participation is guaranteed by \cref{ass:graph}.
In addition, it is important to control the level of asynchrony for the newly introduced communication instants set $T^\ell_{tx}$ in order to bound the maximum delay $D^{\max}$ during the learning. 
This can be done, similar to \cref{ass:asynchrony}, by ensuring that each agent communicates at least every $D>0$ time instants.

Based on \cref{ass:graph}, as well as the introduced communication setup between the agents in \cref{sec:setup:RA}, the following proposition is ensured:
\begin{prop}[maximum delay and global information]\label{ass:buffer:quantity} 
    There exist a finite $D^{\max} \geq D$ such that
    for $t \geq {D^{\max}}$, each agent $i$ has $\bar{r}_\ell^{\mathcal{T}_{i}^{\ell,t}}$ from all agents $\ell\in[m]$ (global information) saved in its local memory, where $t-D^{\max} \leq \mathcal{T}_{i}^{\ell,t} \leq t$.   
\end{prop}
\noindent
\cref{ass:buffer:quantity} emphasizes that although agents may communicate with only their immediate neighbors, they gather \emph{global} information from all the other agents through message passing. It takes $D^{\max}$ time instants for this global information to propagate throughout the graph. Moreover, similar to \cref{prop:globalInfoMaxDel}, the quantities are at most $D^{\max}$ time instants old in each agent's memory.

Since the underlying graph is not fully connected, each agent $i$ receives quantities $\bar{r}_\ell, \forall \ell\in[m]$ with delays as they may pass through intermediate agents who may decide to transmit these quantities only at specific time instants of interest $T_{tx}^j$.
The proposed asynchronous learning approach for DL-based RA is summarized in \cref{alg:proposed:RA}. 
\begin{algorithm}[t]
    \begin{algorithmic}[1]
    \item[\algfont Initialize:] 
        \begin{tabular}[t]{@{}l@{}} 
            $(\gamma_i^t)_{t=0}^{\bar t-1},~ \bm\theta_i^0 \in \R^{n_i}, ~\bar{r}_\ell^0 = 0, \mathcal{T}_{i}^{\ell,0}=0, \ell\in[m]$,\\
            $\mathcal{D}_{u_i} = \mathcal{N}(\bm 0,\bm I), ~\bm u_i^0 \sim \mathcal{D}_{u_i}, \mu > 0, B > 0$,\\
            $T_q^i, T_u^i, T_{tx}^i ~i \in [m]$
        \end{tabular}
    
    \vspace{6pt} 
    \item[\algfont For $t=0,\cdots,\bar t-1$, each agent $i$ asynchronously:]
    \item[\textit{$\blacktriangleright$ Quering:}]
    \vspace{6pt} 
        \State If $t \in T_q^i$: 
            \begin{algsubstates}
                \item compute $\bar{r}_i^t$ using \eqref{eq:barr} with $\bm{\theta}^t_i$ and $\bm u^t_i$
                \item update the memory $\mathcal{B}_i$ by $\{\bar{r}_i^t, ~\text{ID:}~ i, \text{time:}~ t \}$
            \end{algsubstates}
        \vspace{6pt}
        \item[\textit{$\blacktriangleright$ Broadcasting:}]
        \State If $t \in T_{tx}^i$: 
        \begin{algsubstates}
            \item send the quantities in $\mathcal{B}_i$ to the neighbors $j\in\bar{\mathcal{N}}^i_{tx}$
        \end{algsubstates}
        \item[\textit{$\blacktriangleright$ Local memory updates:}]
        \vspace{6pt} 
        \State Receive new updates from the neighbors $j\in\bar{\mathcal{N}}^i_{rx}$ denoted by $Q_\ell=\{r_\ell,\ell,\mathcal{T}_\ell\}, \forall\ell\in[m]$
        \State For each $\ell\in[m]$, if $Q_\ell \neq \emptyset$: 
        \begin{algsubstates}
            \item update the memory $\mathcal{B}_i$ by $\mathcal{T}_{i}^{\ell,t} \gets \mathcal{T}_\ell, ~ \bar{r}_\ell^{\mathcal{T}_{i}^{\ell,t}} \gets r_\ell$
        \end{algsubstates}
        \State $\mathcal{T}_{i}^{\ell,t+1} = \mathcal{T}_{i}^{\ell,t}, \quad \forall \ell\in[m]$
        \State If the memory $\tilde{\mathcal{B}_i}$ is full, omit the oldest entry
        \State $\bm u_i^{t+1} \sim \mathcal{D}_{u_i}$ and save it in the local memory $\tilde{\mathcal{B}_i}$
        \item[\textit{$\blacktriangleright$ Local updates:}]
        \vspace{6pt} 
        \State Update $\bm\theta_i^{t}$ by \eqref{eq:async:individual} and \eqref{eq:s},
        using the locally stored $\bm u_i^{\tau}$ for $\tau \in \{t-D^{\max},\cdots,t\}$ and global information $\bar{r}_\ell^{\mathcal{T}_{i}^{\ell,t}},\forall\ell\in[m]$ 
\end{algorithmic}

% \begin{algorithmic}[1]
%     \item [\algfont Initialize:]
%         \begin{tabular}[t]{@{}l@{}}
%             $\gamma_0 > 0$ and $\tilde\gamma > 0$ for $\gamma^k = \nicefrac{\gamma_0}{(k+1)^{\tilde\gamma}}$,\\ 
%             $B_1>0$, $B_2>0$, $\mu > 0$, $\bm\theta_i^0 \; \forall i \in [N]$
%         \end{tabular}
%     \item[{\algfont each user $i$ asynchronously performs:}]
%     \item[\algfont for {$k=0,1,\dots,$}]
%         \State continuously receive $d_j \; j\neq i$ from the other users
%         \State measure $\E[\bm{H}]{B_1}{U_i}$ % R_i(\bm{H},\bm\phi(\bm{H},\bm{\theta}^k))
%         \State draw a random vector $\bm u_i^k \sim \mathcal{N}(\bm 0, \bm I)$
%         \State update the policy $\phi_i(\cdot,\bm{\theta}^k_i+\mu \bm u_i^k)$
%         \State measure $\E[\bm{H}]{B_2}{U_i}$ % R_i(\bm{H},\bm\phi(\bm{H},\bm{\theta}^k+\mu \bm u^k))
%         \State calculate $d_i^k$ using \eqref{eq:d} and send it to the other users
%         \State perform update on $\bm{\theta}^k_i$ using \eqref{eq:asynchronous_update} \label{alg:s1}
%         \State update the policy $\phi_i(\cdot,\bm{\theta}^{k+1}_i)$
% \end{algorithmic}
    \caption{Proposed asynchronous distributed learning in RA}
    \label{alg:proposed:RA}
\end{algorithm}

The sets $T_q^i, T_{tx}^i$, and $T_u^i$ can be customized by agent $i$ based on factors such as its computational capacity, communication bandwidth, application layer, and other considerations. 
By specifying these sets, agent $i$ has the flexibility to determine when to query its cost, when to communicate with its neighbors, and when to update its local parameters.
It is important to note that the sets $T_q^i, T_{tx}^i$, and $T_u^i$ are not required to be known to agent $i$ before the learning process begins. Rather, they are defined to facilitate the study of the proposed learning approach.
    \section{Numerical Experiments}\label{sec:sim}
        
In the numerical experiments\footnote{The simulation codes are available \href{https://github.com/pourya-b/zoDistLearning.git}{here}.}, our focus is on the application of power allocation in wireless communication networks, where each agent $i$, representing a transmitter, has its own DNN, $\pi_i$, for adjusting its transmit power, \ie $p_i = \pi_i(\bm\theta_i, \bm{\mathcal{H}}_i) \in [0,10]$, where $\bm\theta_i$ is the local parameter vector and $\bm{\mathcal{H}}_i$ represents the input of the DNN defined in the sequel.
The utility of each agent is defined by $r^\pi_i = R_i(\bm p, \bm H)$, where 
\begin{equation}
    R_i(\bm p, \bm H) = \log \left(1+\frac{|h_{r(i)i}|^2 p_i}{1 + \sum_{j\in[m]\setminus i} |h_{r(i)j}|^2 p_j}\right)
    \label{eq:rate}
\end{equation}
represents the achievable data rate of agent $i$, $r(\cdot)$, $h_{ji}$, and $\bm H$ are defined in \cref{sec:system_model}, and $\bm p=(p_1,\cdots,p_m)$ containts the transmit power of all agents. 
The objective of the distributed learning process is to maximize the expected global reward, \ie the sum data rate in the communication network. 
\begin{figure}[t]
    \centering
    \includegraphics[width=0.95\linewidth]{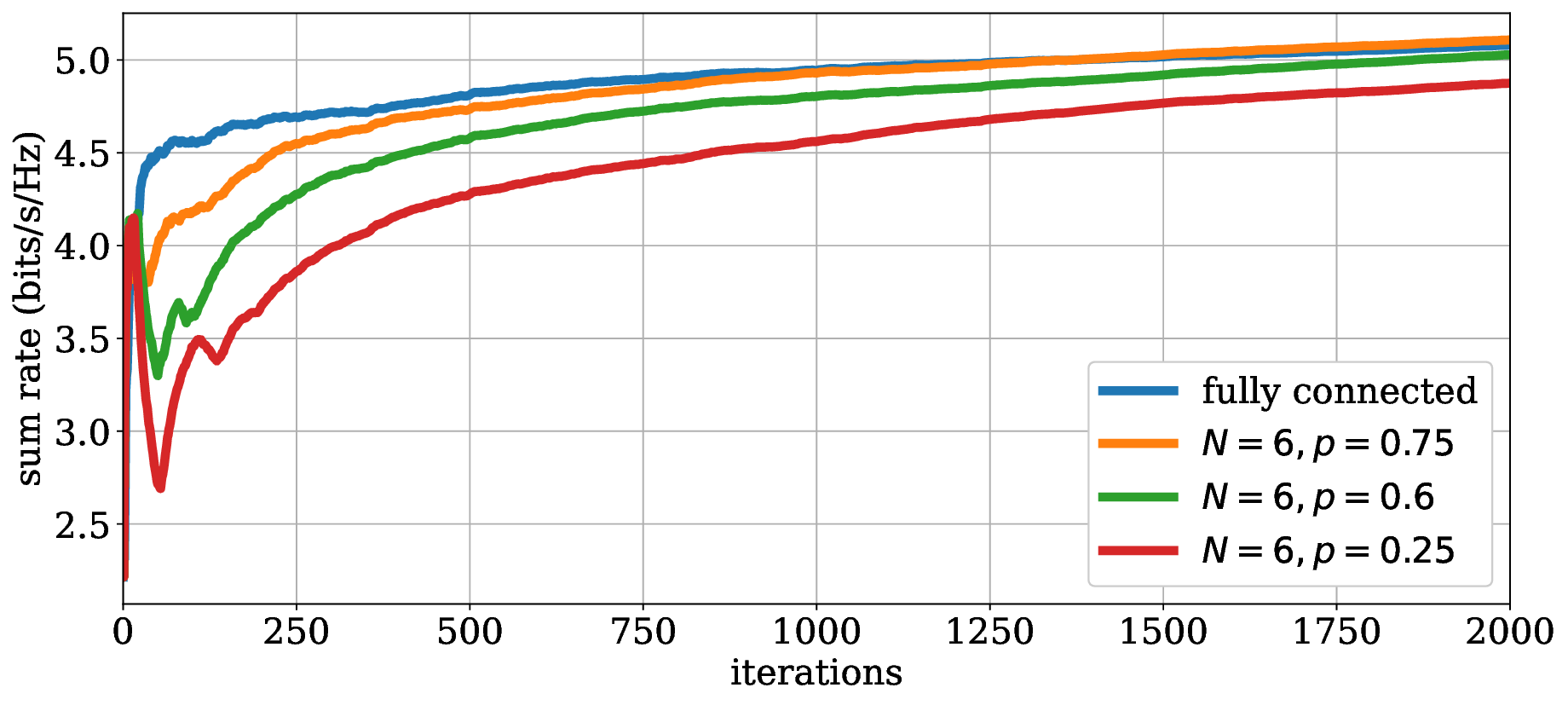}
    \caption{Convergence of the proposed asynchronous distributed learning approach with different values of maximum delay $D^{\rm max}$, defined by \eqref{eq:timestamp} and \cref{ass:buffer:quantity}. There are $m=24$ agents, each communicating with $N$ neighboring agents, and participating in training with probability $p$.
    The comparison is against centralized learning with fully connected agents ($N=m$) in synchronous mode and with no communication delay ($p=1$).}
    \label{fig:Dmax}
\end{figure}%
\newline
\subsubsection*{Wireless communication network setup}
We consider $m$ agents within the communication network with the same number of receivers n=m.
Each agent $i$ is randomly located within the area $l(i) \in [-m, m]^2$. The corresponding receiver $r(i)$ is then randomly located within the area $l(r(i)) \in [l(i) - \nicefrac{m}{4},l(i) + \nicefrac{m}{4}]$.
The considered channel coefficients $h_{r(j)i}$ consist of both large-scale fading, which includes the pathloss gain $h^p_{r(j)i} = \nrm{l(i) - l(r(j))}^{-2.2}, \forall i,j \in [m]$, and small-scale fading represented by $h^f_{r(j)i} \sim \mathcal{C}\mathcal{N}(0,1)$, with a circularly-symmetric complex normal distribution, hence $h_{r(j)i} = h^p_{r(j)i} h^f_{r(j)i}$. Transmitter-receiver pairs that are far from each other may have negligible channel gain, and hence, we exclude them from the policy inputs, defined in the next paragraph. For this, we set $\eta = 0.01$ and use the modified channel $\tilde{\bm{H}}\in\mathbb{C}^{m \times m}$ whose elements are $\tilde{h}_{j i} = h_{r(j)i}$ if $|h_{r(j)i}| \geq \eta$, and $\tilde{h}_{j i} = 0$ otherwise.  
\newline
\subsubsection*{RA setup}
We employ aggregation graph neural networks (GNNs) proposed in \cite{wang_learning_2022}. 
For each policy $i$, the $j$th sample input is the aggregation sequence $\bm{\mathcal{H}}_i^j \coloneqq (y_i^{1,j},\cdots,y_i^{K,j}) \in \R^K$ where $y_i^{k,j}$ is the $i$th element of the vector $\bm y^{k,j} \coloneqq \left[\prod_{j'=0}^{k-1} |\tilde{\bm{H}}^{{j-j'}^T}|\right] \bm 1 \in \R^m$. Here, $K=5$ defines the range of neighbors (hops) that each agent considers in its input, $\bm 1 \in \R^m$ represents a vector with all ones, and $|\cdot|$ is the pointwise absolute value operator.
The RA scheme is distributed, meaning that each agent can compute its policy input locally only through message passing with its neighbors (cf. \cite[eq. (2)]{wang_learning_2022}).

A fully connected DNN with the structure of $\{K, 30, 30, 1\}$ neurons, $\mathrm{relu}(x)=\max\{0,x\}$ activation functions, and sigmoid output function is considered as the policy, hence, $\pi_i(\bm\theta_i,\bm{\mathcal{H}}_i^j) = \mathrm{DNN}(\bm\theta_i, \bm{\mathcal{H}}_i^j)$.
\newline
\subsubsection*{Hyperparameters and the graph $\mathcal{G}^t$}
Consider $A_{m\times m}$ as a matrix consisting of $a_{ij}$, defined in \cref{sec:RAsetup:graph}. 
To ensure that the agents lie on a connected graph $\mathcal{G}^t$ to meet \cref{ass:graph} during the distributed learning, we set $a_{ij}$ such that matrix $A$ is a block diagonal matrix with overlapping blocks. 
These blocks are square matrices with all ones. The size of the square matrices is $\kappa$, with overlaps of $\nicefrac{\kappa}{2}$ elements. Hence, $N \coloneqq |\bar{\mathcal{N}}^i_{rx}| = |\bar{\mathcal{N}}^i_{tx}| = \kappa+\nicefrac{\kappa}{2}, \forall i\in[m]$. 
The communication delay between two directly connected agents is assumed to be one time increment. 
During the numerical experiments, the following parameters in \cref{alg:proposed:RA} are set, unless otherwise specified: $\mu=2, |\mathcal{B}_i|=m, |\tilde{\mathcal{B}_i}|=40$, stepsizes $\gamma_i^t=\frac{0.5}{(t+1)^{0.25}}$, and the batch size $B=20$. 
The other parameters will be specified in the sequel. 
Each agent $i$ decides to update its parameter $\bm\theta_i$ with a probability of $p_u$. Moreover, each agent decides to transmit quantities to each of its neighbors in $\bar{\mathcal{N}}^i_{tx}$ with a probability of $p_{tx}$. 

Centralized learning is considered the benchmark, where individual policies are trained synchronously on a server.
It is noted that this scenario can also be considered as distributed learning where the underlying graph $\mathcal{G}$ is fully connected, there is no communication delay, and the agents are synchronous, \ie $T^i_{tx}=T^i_u=T^i_q=\N$. 
The experiments are repeated 5 times with different random seeds and the average is reported. 
\begin{figure}[t]
    \centering
    \begin{subfigure}{\linewidth}
        \includegraphics[width=0.95\linewidth]{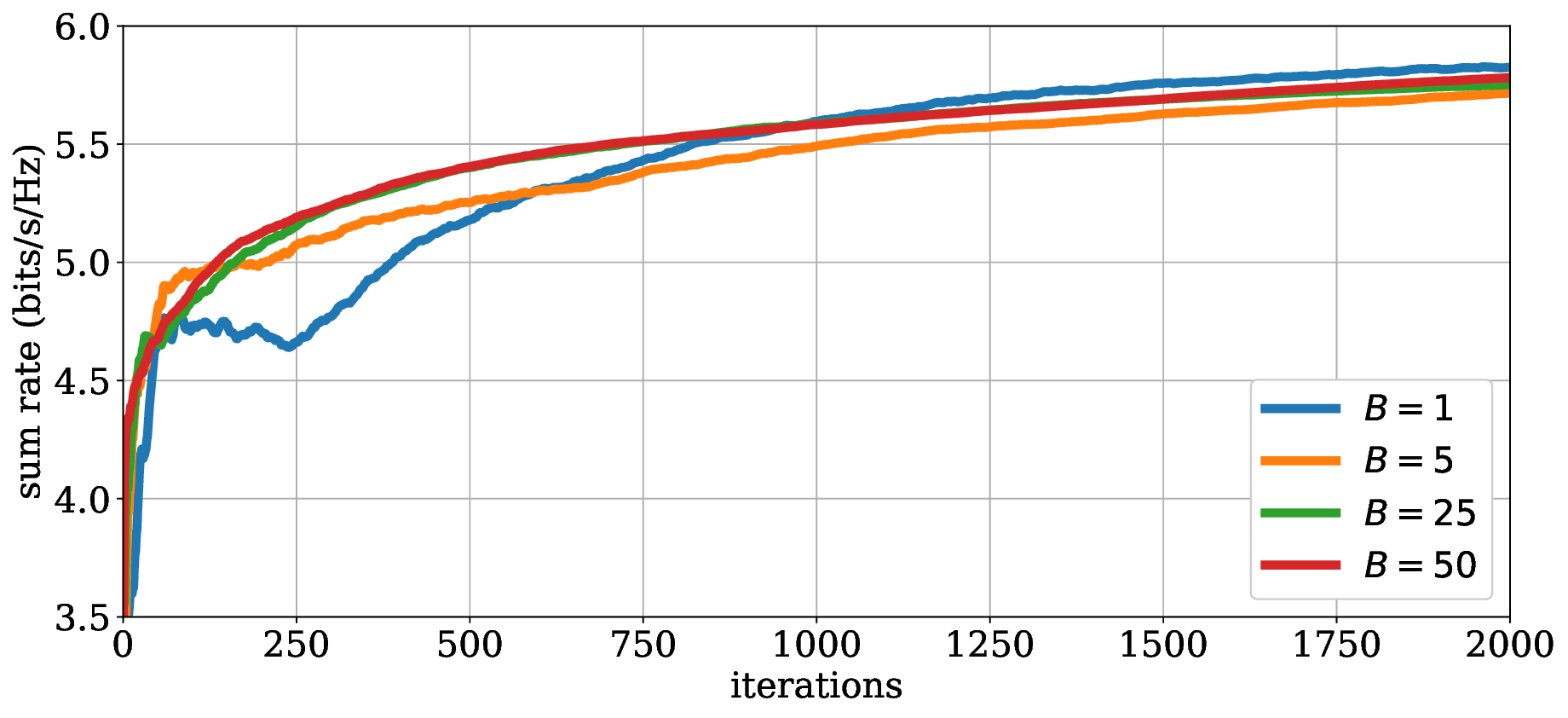}
        \caption{Convergence versus iteration number.}
        \label{fig:batch:1}
    \end{subfigure}
    \begin{subfigure}{\linewidth}
        \includegraphics[width=0.94\linewidth]{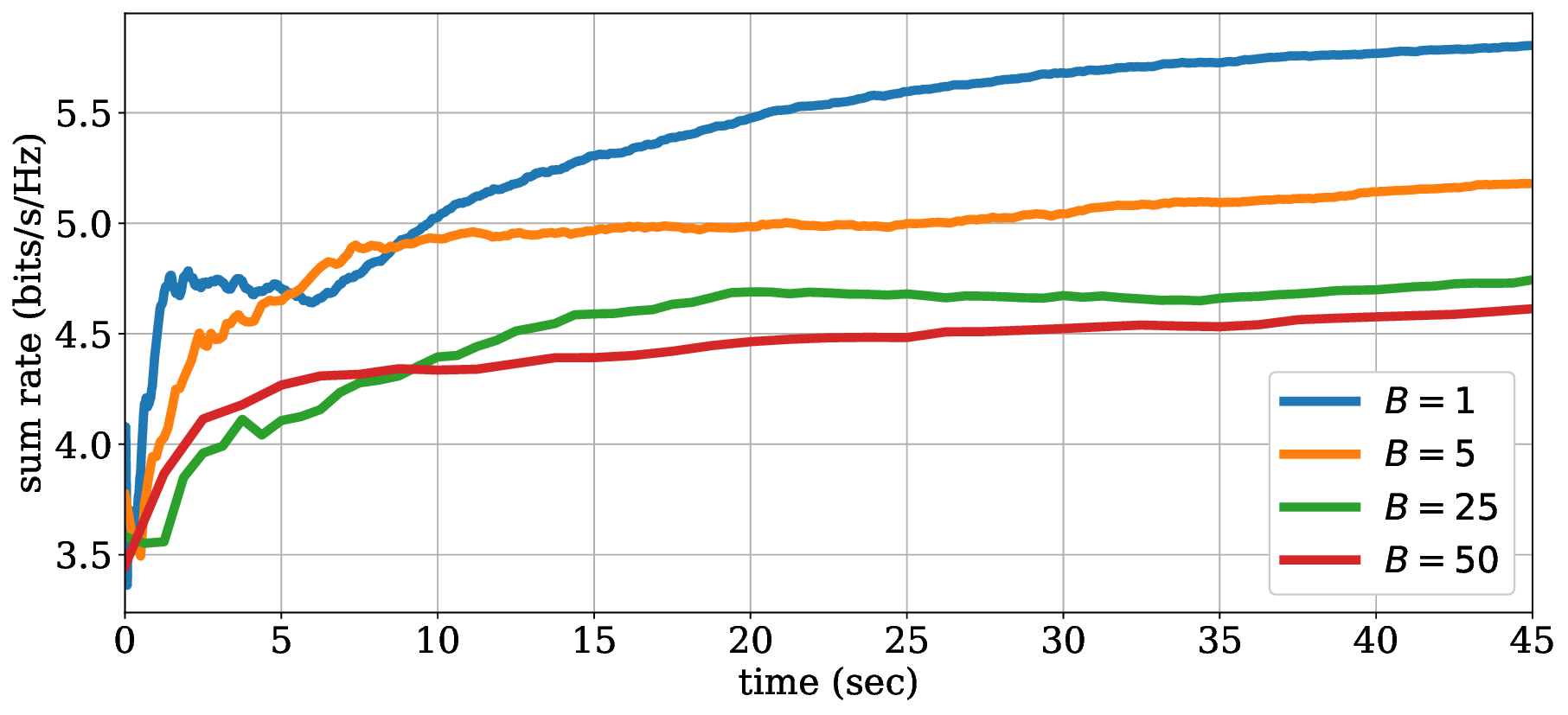}
        \caption{Convergence versus global time.}
        \label{fig:batch:2}
    \end{subfigure}
    \caption{Convergence of the proposed asynchronous distributed learning approach with different batch sizes $B$ defined in \eqref{eq:barr}. There are $m = 24$ agents, each communicating with $N=4$ neighboring agents, and participating in training with a probability of $p=0.9$. 
    The duration of one iteration in (b) is set equal to the typical average channel coherence time of 25 ms.}
    \label{fig:batch}
\end{figure}

\bigskip
\subsubsection*{Results}

In the first experiment, we assess the performance of the proposed learning approach under varying probabilities $p \coloneqq p_u =p_{tx}$, with $m=24$ agents. 
This parameter, along with the number of neighboring agents $N$, directly impacts the maximum delay $D^{\max}$ in \cref{thm:subseq} which determines the maximum age of the stored quantities $\bar{r}_i$ in each agent's memory. A lower probability $p$ results in a longer propagation time for the quantities $\bar{r}_i$ throughout the network (refer to \cref{sec:alg:RA} for further explanations). 
The observations in \cref{fig:Dmax}, where the performance is reported versus iteration number $t$, are aligned with the theoretical findings in \cref{thm:subseq}, as increasing $D^{\max}$ leads to slower convergence. In addition, convergence is comparable with centralized learning, even with probabilities as low as $p=0.25$.
\begin{figure}[t]
    \centering
    \includegraphics[width=0.95\linewidth]{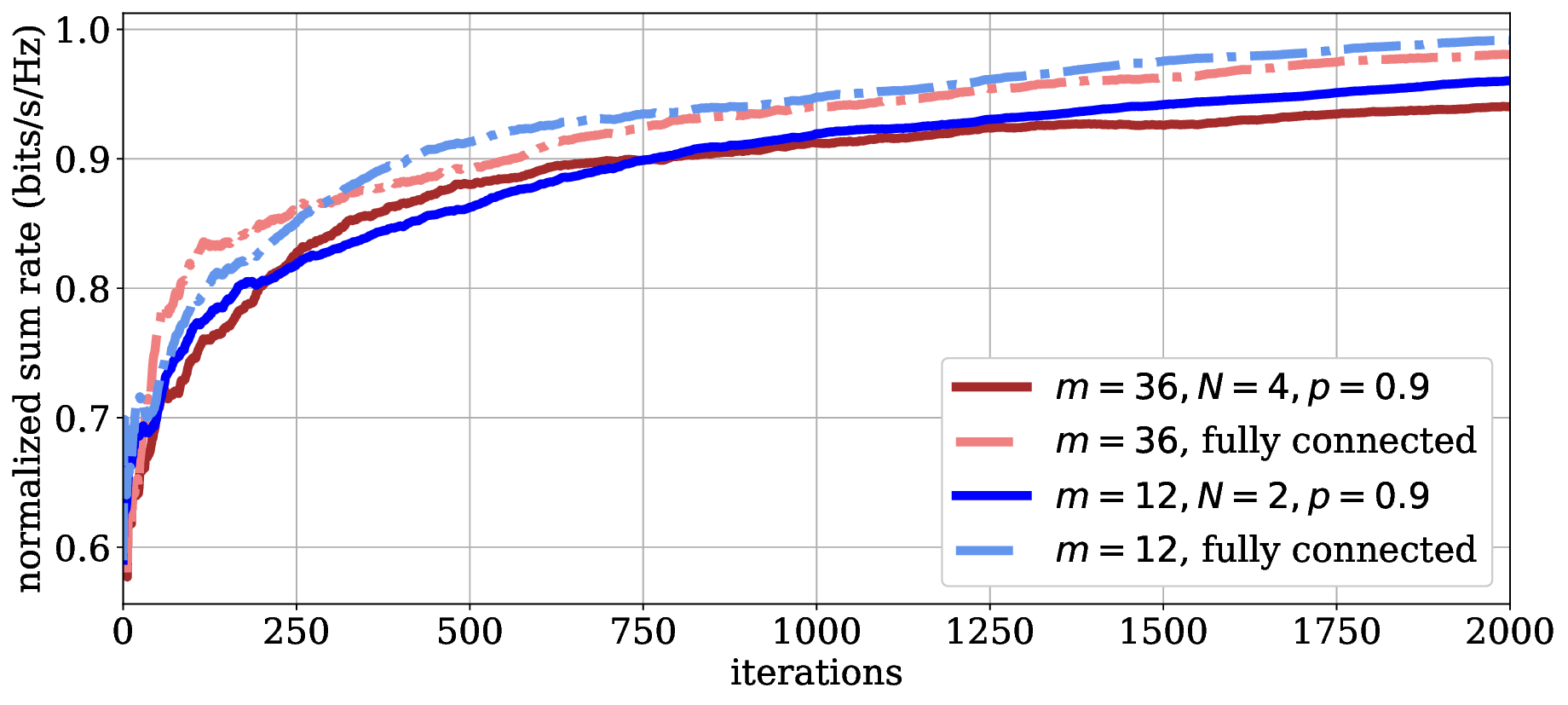}
    \caption{Convergence of the proposed asynchronous distributed learning approach with different numbers of agents $m$.
    Each agent communicates with $N$ neighboring agents and participates in training with probability $p$. 
    The performance is normalized by the maximum achieved sum rate for each $m$.
    The comparison is against centralized learning with fully connected agents ($N=m$) in synchronous mode and with no communication delay ($p=1$).}
    \label{fig:agent_num}
\end{figure}

\cref{fig:batch:1} presents the impact of batch size on the convergence rate. The findings are consistent with theoretical results, as larger batch sizes lead to faster convergence. 
However, it is important to note that agents require more time to accumulate a larger number of samples for larger batch sizes. 
As a result, the overall convergence becomes slower over time for larger batch sizes. This phenomenon is demonstrated in \cref{fig:batch:2}, where the duration of one increment of global time is set equal to the typical average channel coherence time of $25$ ms in wireless communication networks.

The convergence of the proposed approach is also depicted for different numbers of agents, specifically $m=12$ and $m=36$ in \cref{fig:agent_num}.

    \section{Conclusion}
        In this paper, we have investigated the problem of distributed learning where agents collaborate on a common task while they also have distinct individual tasks.
In this context, the performance of each agent is influenced by other agents through interactions. 
To optimize their local parameters, agents utilize zeroth-order oracles and exchange their local queries through a message-passing architecture, subject to random but bounded communication delays. 
Importantly, the shared quantities are scalar, ensuring efficient communication bandwidth usage and preserving agents' privacy.
Moreover, the agents engage in distributed learning in an asynchronous manner. 
We have also established convergence analyses for smooth nonconvex problems, achieving the same convergence rate as existing learning approaches with synchronous updates. According to the analysis results, the stepsize needs to be decreased for larger delays.

As a use-case, the relevant problem of DL-based RA in communication networks is addressed, where agents, acting as transmitters, collaborate in training their local DNNs using the proposed learning approach to maximize the expected global reward in the communication network. 
An interesting direction for future research is to extend the proposed learning approach to nonsmooth regularized settings and investigate its application in other learning tasks.
 
\appendices
    \section{Preliminaries}\label{sec:app:pre}
        
\begin{table}[t]
    {
    \centering
    \begin{tabular}{|c|c||c|c|}
        \hline
        not. & Description & not. & Description\\
        \hline
        $m$ & number of agents & $\gamma_\ell$ & stepsize\\
        $f_\ell$ & local cost function & $T_q^\ell$ & set of query instants\\
        $\bm\theta_\ell$ & local parameter vector & $T_u^\ell$ & set of update instants\\
        $F_\ell$ & local cost value & $\mathcal{B}_\ell, \tilde{\mathcal{B}}_\ell$ & local memories\\
        $f^\mu_\ell$ & smooth approx. function & $B$ & batch size\\
        $\tilde{f}_\ell$ & surrogate local function & $\Theta_\ell$ & effective parameters\\
        $\mathcal{N}^{\ell}_{rx}$ & neighbor set & $\tilde{\bm s}_\ell$ & estimated gradient \\
        $\mathcal{N}^{\ell}_{tx}$ & neighbor set & $\bm u_\ell$ & perturbation vector \\
        $\bar{r}_\ell$ & local queries & $D^{\rm max}$ & maximum delay time\\
        $\mu$ & smoothing parameter & $\bm{\xi}$ & sample vector\\
        \hline
    \end{tabular}
    \caption{List of notations}
    \label{tab:notations}
    }
\end{table}

Frequently used notations throughout the paper are listed in \cref{tab:notations}.
Define the following filtration:
\begin{equation}
  \mathcal{F}^t \coloneqq \filt \{\bm{\xi}^k, \bm u^k \mid k\in[t-1]\}.
  \label{eq:rnd}
\end{equation}
The filtration $\mathcal{F}^t$ includes all the randomness involved up to the time instant $t$.
During the proof, we make use of the following Young's inequality for any $a, b \geq 0$,
\begin{equation}
  ab \leq \nicefrac{a^2}{2} + \nicefrac{b^2}{2},
  \label{eq:Young}
\end{equation}
and the following lemma and propositions:

\begin{lem}[{descent lemma \cite[Prop. A.24]{bertsekas1997nonlinear}}]\label{lem:descent}
  Given a continuous and $L_g$-smooth function $\func{g}{\R^n}{\R}$, the following holds for all $\bm{\theta}',\bm{\theta} \in \R^n$:
  \begin{align*}
    g(\bm{\theta}') \leq g(\bm{\theta}) + \langle \nabla g(\bm{\theta}), \bm{\theta}' - \bm{\theta} \rangle + \frac{L_g}{2} \sqnrm{\bm{\theta}'-\bm{\theta}}.
  \end{align*}
\end{lem}

\begin{prop}[Lipschitz continuous and differentiable $\tilde{f}_\ell$]\label{prop:lipschitz}
  The function $\tilde{f}_\ell$ defined in \eqref{eq:s_expect} is Lipschitz continuous and $\tilde L$-smooth with some constants $\tilde L^0 \in [0,\nicefrac{L^0}{\sqrt{m}}]$ and $\tilde L \in [0,\nicefrac{L}{\sqrt{m}}]$, where $L^0$ and $L$ are defined in \cref{ass:lipschitz}. 
  Namely, take $\Theta_\ell,\Theta'_\ell \in \R^{n|\mathcal{N}^{\ell}_{rx}|}$, then
  \begin{equation*}
    \nrm{\nabla_\ell \tilde{f}_\ell(\Theta_\ell) - \nabla_\ell \tilde{f}_\ell(\Theta_\ell')} \leq \tilde L \nrm{\Theta_\ell - \Theta_\ell'},~ \forall \Theta_\ell,\Theta_\ell'\in \R^{n|\mathcal{N}^{\ell}_{rx}|}.
  \end{equation*}
  \begin{proof}
    By \eqref{eq:s_expect}, we rewrite the lhs as:
    \begin{align*}
      &\sqnrm{\frac{1}{m} \sum_{i\in\mathcal{N}^{\ell}_{rx}} \left(\nabla_\ell f_{i}(\bm{\theta}^i) - \nabla_\ell f_{i}(\bm{\theta}^{\prime^i})\right)}\\
      &\leq  \frac{1}{m} \sum_{i\in\mathcal{N}^{\ell}_{rx}} \sqnrm{\nabla_\ell f_{i}(\bm{\theta}^i) - \nabla_\ell f_{i}(\bm{\theta}^{\prime^i})}\\
      &\leq \frac{L^2}{m} \sum_{i\in\mathcal{N}^{\ell}_{rx}} \sqnrm{\bm{\theta}^i - \bm{\theta}^{\prime^i}}
      = \frac{L^2}{m} \sqnrm{\Theta_\ell - \Theta_\ell'},
    \end{align*}
    where $\bm{\theta}^i \in\R^n$ and $\bm{\theta}^{\prime^i}\in\R^n$ are $i$th block coordinates of $\Theta_\ell$ and $\Theta'_\ell$, respectively.
    The first inequality is due to Jensen's inequality, the second is due to \cref{ass:lipschitz:1}, and the last equality is due to the definition of $\Theta_\ell$ and $\Theta'_\ell$. The same reasoning can be leveraged for the Lipschitz constant $\tilde L^0$.
  \end{proof}
\end{prop}

\begin{prop}[Lipschitz differentiable $f^\mu_\ell$ and $\tilde{f}^\mu_\ell$] \label{lem:grad_distance}
  For all $\ell \in [m]$, the functions $f^\mu_\ell$ defined in \eqref{eq:s_expect} are $L_\mu$-smooth with $L_\mu \leq L$. 
  Moreover, take $\Theta^t_\ell$ and $\bar\Theta_\ell^t$ defined in \eqref{eq:Params} and \eqref{eq:theta_common}, respectively.  
  The following is true for the function $\tilde{f}^\mu_\ell$ in \eqref{eq:s_expect}:
  \begin{align*}
    \Vert \nabla_\ell \tilde{f}^\mu_\ell(\bar\Theta^t_\ell) &- \nabla_\ell \tilde{f}^\mu_\ell(\Theta^t_\ell) \Vert\\
    &\leq \frac{L}{m} \sum_{i\in \mathcal{N}^\ell_{rx}} \sum_{k \in [m]}  \nrm{\bm{\theta}^t_k - \bm{\theta}^{\mathcal{T}_{\ell}^{i,t}}_k}, \quad \forall\ell\in[m]
  \end{align*}
  \begin{proof}
    The $L_\mu$-smoothness of the functions $f^\mu_\ell$ can be verified by \cref{ass:lipschitz:1,prop:approximation:1}. For the inequality, consider
    \begin{align*}
      \Vert \nabla_\ell \tilde{f}^\mu_\ell(\bar\Theta^t_\ell &) - \nabla_\ell \tilde{f}^\mu_\ell(\Theta^t_\ell) \Vert \\
      &\leq \frac{1}{m} \sum_{i\in\mathcal{N}^\ell_{rx}} \nrm{\nabla_\ell f^\mu_{i}(\bm{\theta}^t) - \nabla_\ell f^\mu_{i}(\bm{\theta}^{\mathcal{T}_{\ell}^{i,t}})}\\
      & \leq \frac{L}{m} \sum_{i\in\mathcal{N}^\ell_{rx}} \nrm{\bm{\theta}^t - \bm{\theta}^{\mathcal{T}_{\ell}^{i,t}}},
    \end{align*}
    where the first inequality uses the definition of the function in \eqref{eq:s_expect}, and the second inequality considers the $L_\mu$-smoothness of the involved functions and $L_\mu \leq L$ guaranteed by \cref{prop:approximation:1}.
  \end{proof}
\end{prop}

\begin{prop}[unbiased oracle $\bm{\tilde{s}}^t_\ell$]\label{lem:unbiased}
  For all $\ell\in[m]$ and $t\in T_u^\ell$, $\bm{\tilde{s}}^t_\ell$ defined in \eqref{eq:s} is unbiased:
  \[
      \E[\bm{\xi},\mathbb{U}]{\bm{\tilde{s}}^t_\ell} = -\nabla_\ell \tilde{f}^\mu_\ell(\Theta_\ell^t),
  \]
  where $\tilde{f}^\mu_\ell$ is defined in \eqref{eq:s_expect}.
  \begin{proof}
    Based on the timestamp $\mathcal{T}_{\ell}^{i,t}$ stored in $\mathcal{B}_\ell$, user $\ell$ can determine that the vector $\bm u_\ell^{\mathcal{T}_{\ell}^{i,t}}$ stored in $\tilde{\mathcal{B}_{\ell}}$ was used during the measurement of $\bar{r}_i^{\mathcal{T}_{\ell}^{i,t}}$ by user $i$. This inference is facilitated by \cref{ass:timing,ass:buffer:size}. Consequently, user $\ell$ can establish \eqref{eq:s} by leveraging the definition of $\bar{r}_i$ in \eqref{eq:barr}, \cref{prop:approximation:2}, and the unbiased nature of the oracle $F_i$ as indicated by \cref{ass:lipschitz:unbiased}. This series of reasoning allows for the derivation of the equality.
  \end{proof}
\end{prop}

\begin{prop}[oracle $\bm{\tilde{s}}^t_\ell$ with bounded variance]\label{prop:BoundedOracle}
  For all $\ell\in[m]$ and $t\in T_u^\ell$, $\bm{\tilde{s}}^t_\ell$ defined in \eqref{eq:s} is bounded:
    \[
        \E[\bm{\xi},\mathbb{U}]{\sqnrm{\bm{\tilde{s}}^t_\ell + \nabla_\ell \tilde{f}^\mu_\ell(\Theta_\ell^t)}} \leq \nicefrac{\tilde\sigma^2}{B}, 
    \]
    where $B$ is the batchsize used by all the users in \eqref{eq:barr} to derive $\bm{\tilde{s}}^t_\ell$ and
    \[
        \tilde\sigma^2 \coloneqq 4(n+4) (\sigma^2 + G^2) + {\mu^2}L^2(n+6)^3 + 4\frac{B L^2_{\bm{\xi}}}{\mu^2} v^2.
    \]
    \begin{proof}
      The proof follows the proof in \cite[thm. 4]{ghadimi_mini-batch_2016} with necessary modifications.
      Considering \eqref{eq:s} and \eqref{eq:s_expect} and $\mathcal{T}$ instead of $\mathcal{T}_{\ell}^{i,t}$ for the sake of presentation, we have
      \begin{align*}
        & \E[\bm{\xi},\mathbb{U}]{\sqnrm{\bm{\tilde{s}}^t_\ell + \nabla_\ell \tilde{f}^\mu_\ell(\Theta_\ell^t)}}\\
        & = \E[\bm{\xi},\mathbb{U}]{\sqnrm{\sum_{i\in\mathcal{N}^\ell_{rx}} \left(\bar{r}_i^{\mathcal{T}} \bm u_\ell^{\mathcal{T}} + \frac{1}{m}\nabla_\ell f^\mu_i(\bm{\theta}^{\mathcal{T}})\right)}}\\
        &\leq m \sum_{i\in\mathcal{N}^\ell_{rx}} \E[\bm{\xi},\mathbb{U}]{\sqnrm{\bar{r}_i^{\mathcal{T}} \bm u_\ell^{\mathcal{T}} + \frac{1}{m}\nabla_\ell f^\mu_i(\bm{\theta}^{\mathcal{T}})}}\\
        &=\frac{1}{m} \sum_{i\in\mathcal{N}^\ell_{rx}} \\
        &\mathbb{E}_{\bm{\xi},\mathbb{U}} \bigg[ \bigg\| \frac{\frac{1}{B}\sum_{q=1}^{B} \left(F_i(\bm{\theta}^{\mathcal{T}},\bm{\xi}^q) - F_i(\bm{\theta}^{\mathcal{T}}+\mu \bm u^{\mathcal{T}},\bm{\xi}^q)\right)}{\mu}\bm u_\ell^{\mathcal{T}} \\
        &+ \nabla_\ell f^\mu_i(\bm{\theta}^{\mathcal{T}}) + \frac{\frac{1}{B}\sum_{q=1}^{B} \left(F_i(\bm{\theta}^{\mathcal{T}},\bm{\xi}^{\prime^q}) - F_i(\bm{\theta}^{\mathcal{T}},\bm{\xi}^q)\right)}{\mu}\bm u_\ell^{\mathcal{T}}  \bigg\|^2 \bigg]\\
        &\leq \frac{2}{B^2 m} \sum_{i\in\mathcal{N}^\ell_{rx}} \sum_{q=1}^{B} \\
        &\mathbb{E}_{\bm{\xi},\mathbb{U}} \bigg[ \bigg\| \frac{F_i(\bm{\theta}^{\mathcal{T}},\bm{\xi}^q) - F_i(\bm{\theta}^{\mathcal{T}}+\mu \bm u^{\mathcal{T}},\bm{\xi}^q)}{\mu}\bm u_\ell^{\mathcal{T}} + \nabla_\ell f^\mu_i(\bm{\theta}^{\mathcal{T}}) \bigg\|^2 \bigg]\\
        & + \frac{2}{\mu^2 B m} \sum_{i\in\mathcal{N}^\ell_{rx}} 
        \sum_{q=1}^{B} \\
        &\E[\bm{\xi}]{\sqnrm{F_i(\bm{\theta}^{\mathcal{T}},\bm{\xi}^q) - F_i(\bm{\theta}^{\mathcal{T}},\bm{\xi}^{\prime^q})}}
        \E[\mathbb{U}]{\sqnrm{\bm u_\ell^{\mathcal{T}}}} \numberthis \label{eq:VarDerivation}
      \end{align*}
      where in the first inequality Jensen's inequality is applied with taking into account $|\mathcal{N}^\ell_{rx}| \leq m$. In the second equality, $\bar{r}_i$ is expanded using \eqref{eq:barr}, where the batch size is $B$.
      Also, $F_i(\bm{\theta}^\mathcal{T},\bm{\xi}^q)$ is added and subtracted.
      For the last inequality, Jensen's inequality is applied for both terms, while for the first term \cref{prop:approximation:2} is also considered, with $\E{\sqnrm{\frac{1}{B}\sum_{q=1}^B \nabla H(\bm\theta,\bm\xi^q) - \E{\nabla H(\bm\theta,\bm\xi)}}} = \frac{1}{B^2}\sum_{q=1}^B \E{\sqnrm{\nabla H(\bm\theta,\bm\xi^q) - \E{\nabla H(\bm\theta,\bm\xi)}}}$, a function $H$, and sample vectors $\bm\xi^q$ and $\bm\xi$.
      We continue to bound the first term in the last inequality:
      \begin{align*}
        & \E[\bm{\xi},\mathbb{U}]{\sqnrm{\frac{F_i(\bm{\theta}^{\mathcal{T}},\bm{\xi}^q) - F_i(\bm{\theta}^{\mathcal{T}}+\mu \bm u^{\mathcal{T}},\bm{\xi}^q)}{\mu}\bm u_\ell^{\mathcal{T}} + \nabla_\ell f^\mu_i(\bm{\theta}^{\mathcal{T}})}} \\
        & \leq \E[\bm{\xi},\mathbb{U}]{\sqnrm{\frac{F_i(\bm{\theta}^{\mathcal{T}}+\mu \bm u^{\mathcal{T}},\bm{\xi}^q) - F_i(\bm{\theta}^{\mathcal{T}},\bm{\xi}^q)}{\mu}\bm u_\ell^{\mathcal{T}} }} \\
        &\leq 2(n+4)\E[\bm{\xi}]{\sqnrm{\nabla F_i(\bm{\theta}^{\mathcal{T}},\bm{\xi}^q) - \nabla f_i(\bm{\theta}^{\mathcal{T}}) + \nabla f_i(\bm{\theta}^{\mathcal{T}})}} \\
        & + \frac{\mu^2}{2}L^2(n+6)^3\\
        &\leq 2(n+4) (\sigma^2 + G^2) + \frac{\mu^2}{2}L^2(n+6)^3,
      \end{align*}
      where in the first inequality $\nabla_\ell f^\mu_i(\bm{\theta}^\mathcal{T})$ is omitted by the fact that the variance is upperbounded by the second moment. We use \cref{prop:approximation:3} in the second inequality, and \cref{ass:oracle:1,ass:oracle:2} in the last inequality.
      The last term in \eqref{eq:VarDerivation} is also bounded by
      \begin{align*}
        \E[\bm{\xi}]{\sqnrm{F_i(\bm{\theta}^{\mathcal{T}},\bm{\xi}^q) - F_i(\bm{\theta}^{\mathcal{T}},\bm{\xi}^{\prime^q})}}
        &\leq L_{\bm{\xi}}^2 \E[\bm{\xi}]{\sqnrm{\bm{\xi}^q - \bm{\xi}^{\prime^q}}} \\
        &\leq 2L_{\bm{\xi}}^2 v^2,
      \end{align*}
      leveraging \cref{ass:lipschitz:3,ass:oracle:3}. 
      Considering $\E{\sqnrm{\bm u_\ell^{\mathcal{T}}}} \leq 1$ due to its distribution and putting the last two bounds back into \eqref{eq:VarDerivation}, where  $|\mathcal{N}^\ell_{rx}| \leq m$, completes the proof.
    \end{proof}
\end{prop}
    \section{Convergence study}\label{sec:app:conv}
        \begin{appendixproof}{thm:subseq}\label{thm:subseq:proof}
    We make use of the descent lemma (refer to \cref{lem:descent}) on the $L_\mu$-smooth function ${f}^{\mu}(\bm{\theta}^t)$ with an expectation conditioned on the filtration $\mathcal{F}^t$, defined in \eqref{eq:rnd}. Note that as conditioned on $\mathcal{F}^t$, all the iterates up to and including $\bm{\theta}^t$ are deterministic:
    \begin{align*}
            &\E{{f}^{\mu}(\bm{\theta}^{t+1})}[\mathcal{F}^t] \leq {f}^{\mu}(\bm{\theta}^t) + \E{\langle \nabla {f}^{\mu}(\bm{\theta}^t), \bm{\theta}^{t+1} - \bm{\theta}^t \rangle}[\mathcal{F}^t] \\
            & + \frac{L_\mu}{2}\E{\sqnrm{\bm{\theta}^{t+1}-\bm{\theta}^t}}[\mathcal{F}^t] \label{eq:descent_lemma}\numberthis\\
            & \leq {f}^{\mu}(\bm{\theta}^t) + \sum_{\ell=1}^m \E{\langle \nabla_\ell{f}^{\mu}(\bm{\theta}^t), \gamma^t_\ell \bm{\tilde{s}}_\ell^t \rangle}[\mathcal{F}^t]\\ 
            &+ \frac{L}{2}\sum_{\ell=1}^m (\gamma_\ell^t)^2 \E{\sqnrm{\bm{\tilde{s}}_\ell^t}}[\mathcal{F}^t]\\
            & = {f}^{\mu}(\bm{\theta}^t) + \sum_{\ell=1}^m \E{\langle \nabla_\ell\tilde{f}^\mu_\ell(\bar\Theta^t_\ell), \gamma^t_\ell \bm{\tilde{s}}_\ell^t \rangle}[\mathcal{F}^t]\\
            &+ \frac{L}{2}\sum_{\ell=1}^m (\gamma_\ell^t)^2 \E{\sqnrm{\bm{\tilde{s}}_\ell^t}}[\mathcal{F}^t]\\
            & = {f}^{\mu}(\bm{\theta}^t) + \sum_{\ell=1}^m \E{\langle \nabla_\ell\tilde{f}^\mu_\ell(\Theta^t_\ell), \gamma^t_\ell \bm{\tilde{s}}_\ell^t \rangle}[\mathcal{F}^t]\\
            &+ \frac{L}{2}\sum_{\ell=1}^m (\gamma_\ell^t)^2 \E{\sqnrm{\bm{\tilde{s}}_\ell^t}}[\mathcal{F}^t] \\
            & + \sum_{\ell=1}^m \E{\langle \nabla_\ell\tilde{f}^\mu_\ell(\bar\Theta^t_\ell)-\nabla_\ell\tilde{f}^\mu_\ell(\Theta^t_\ell), \gamma^t_\ell \bm{\tilde{s}}_\ell^t \rangle}[\mathcal{F}^t], 
    \end{align*}
    where in the second inequality $L_\mu \leq L$ is considered by \cref{prop:approximation:1} and \eqref{eq:update_rule} is employed, in the first equality \eqref{eq:funcs:theta_common} is used.
    We proceed by bounding the last term in \eqref{eq:descent_lemma} as
    \begin{align*}
        &\sum_{\ell=1}^m \E{\langle \nabla_\ell\tilde{f}^\mu_\ell(\bar\Theta^t_\ell)-\nabla_\ell\tilde{f}^\mu_\ell(\Theta^t_\ell), \gamma^t_\ell \bm{\tilde{s}}_\ell^t \rangle}[\mathcal{F}^t] \label{eq:descent_lemma:lastterm} \numberthis\\
        &\leq \sum_{\ell=1}^m \nrm{\nabla_\ell\tilde{f}^\mu_\ell(\bar\Theta^t_\ell)-\nabla_\ell\tilde{f}^\mu_\ell(\Theta^t_\ell)} \E{\nrm{\gamma^t_\ell \bm{\tilde{s}}_\ell^t}}[\mathcal{F}^t]\\
        &\leq \frac{L}{m} \sum_{\ell=1}^m \sum_{i\in\mathcal{N}^{\ell}_{rx}}  \sum_{k=1}^m \nrm{\bm{\theta}^t_k - \bm{\theta}^{\mathcal{T}_{\ell}^{i,t}}_k} \E{\nrm{\gamma^t_\ell \bm{\tilde{s}}_\ell^t}}[\mathcal{F}^t]\\
        & \leq \frac{L}{m} \sum_{\ell=1}^m \sum_{i\in\mathcal{N}^{\ell}_{rx}}  \sum_{k=1}^m \sum_{\tau=\mathcal{T}_{\ell}^{i,t}}^{t-1} \E{\gamma^\tau_k \gamma^t_\ell\nrm{\bm{\tilde{s}}_k^\tau}\nrm{\bm{\tilde{s}}_\ell^t}}[\mathcal{F}^t]\\
        & \leq \frac{L}{m} \sum_{\ell=1}^m \sum_{i\in\mathcal{N}^{\ell}_{rx}}  \sum_{k=1}^m \sum_{\tau=t-D^{\max}}^{t-1} \\
        &\E{\frac{(\gamma^\tau_k)^2 \sqnrm{\bm{\tilde{s}}_k^\tau}}{2} +  \frac{(\gamma^t_\ell)^2\sqnrm{\bm{\tilde{s}}_\ell^t}}{2}}[\mathcal{F}^t]\\
        & \leq \frac{m L}{2} \sum_{\ell=1}^m \sum_{\tau=t-D^{\max}}^{t-1} \E{{(\gamma^\tau_\ell)^2 \sqnrm{\bm{\tilde{s}}_\ell^\tau}} +  {(\gamma^t_\ell)^2\sqnrm{\bm{\tilde{s}}_\ell^t}}}[\mathcal{F}^t],
    \end{align*}
    where the second inequality is due to \cref{lem:grad_distance}, in the third inequality \eqref{eq:async:individual} is employed, and in the fourth inequality, Young's inequality in \eqref{eq:Young} and \cref{ass:asynchrony:2} are considered.
    The last inequality is also due to the bound $|\mathcal{N}^{\ell}_{rx}|\leq m$.
    Putting \eqref{eq:descent_lemma:lastterm} into \eqref{eq:descent_lemma} results in 
    \begin{align*}
        &\E{{f}^{\mu}(\bm{\theta}^{t+1})}[\mathcal{F}^t] \label{eq:descent:pre}\numberthis\\
        &\leq {f}^{\mu}(\bm{\theta}^t) +\sum_{\ell=1}^m \E{\langle \nabla_\ell\tilde{f}^\mu_\ell(\Theta^t_\ell), \gamma^t_\ell \bm{\tilde{s}}_\ell^t \rangle}[\mathcal{F}^t] \\
        &+ L\frac{m D^{\max} + 1}{2} \sum_{\ell=1}^m (\gamma^t_\ell)^2 \E{\sqnrm{\bm{\tilde{s}}_\ell^t}}[\mathcal{F}^t]\\
        &+ \frac{m L}{2} \sum_{\ell=1}^m\sum_{\tau=t-D^{\max}}^{t-1} (\gamma^\tau_\ell)^2 \sqnrm{\bm{\tilde{s}}_\ell^\tau}.
    \end{align*}
    By \cref{lem:unbiased}:
    \begin{align*}
        \sum_{\ell=1}^m \E{\langle \nabla_\ell\tilde{f}^\mu_\ell(\Theta^t_\ell), \gamma^t_\ell \bm{\tilde{s}}_\ell^t \rangle}[\mathcal{F}^t] 
        = -\sum_{\ell=1}^m \gamma^t_\ell \sqnrm{\nabla_\ell\tilde{f}^\mu_\ell(\Theta^t_\ell)}
    \end{align*}
    Moreover, by \cref{lem:unbiased,prop:BoundedOracle}, 
    \begin{align*}
        \E{\sqnrm{\bm{\tilde{s}}_\ell^t}}[\mathcal{F}^t] 
        &= \E{\sqnrm{\bm{\tilde{s}}_\ell^t+ \nabla_\ell\tilde{f}^\mu_\ell(\Theta^t_\ell) - \nabla_\ell\tilde{f}^\mu_\ell(\Theta^t_\ell)} }[\mathcal{F}^t]\\
        &\leq \nicefrac{\tilde\sigma^2}{B} + \sqnrm{\nabla_\ell\tilde{f}^\mu_\ell(\Theta^t_\ell)}. \label{eq:mean_s}\numberthis
    \end{align*}
    Taking the total expectation of \eqref{eq:descent:pre} by applying the tower rule and then utilizing the above relations lead to:
    \begin{align*}
        &\E{{f}^{\mu}(\bm{\theta}^{t+1})}
        \leq \E{{f}^{\mu}(\bm{\theta}^t)} -\sum_{\ell=1}^m \gamma^t_\ell \E{\sqnrm{\nabla_\ell\tilde{f}^\mu_\ell(\Theta^t_\ell)}}\\
        &+ L\frac{m D^{\max} + 1}{2} \sum_{\ell=1}^m (\gamma^t_\ell)^2 \E{\sqnrm{\bm{\tilde{s}}_\ell^t}}\\
        &+ L\frac{m}{2} \sum_{\ell=1}^m\sum_{\tau=t-D^{\max}}^{t-1} (\gamma^\tau_\ell)^2 \E{\sqnrm{\bm{\tilde{s}}_\ell^\tau}}\\
        &\leq \E{{f}^{\mu}(\bm{\theta}^t)} - \sum_{\ell=1}^m \gamma^t_\ell \E{\sqnrm{\nabla_\ell\tilde{f}^\mu_\ell(\Theta^t_\ell)}}\\
        &+ L\frac{m D^{\max} + 1}{2} \sum_{\ell=1}^m (\gamma^t_\ell)^2 \E{\sqnrm{\nabla_\ell\tilde{f}^\mu_\ell(\Theta^t_\ell)}}\\
        &+ L\frac{m}{2} \sum_{\ell=1}^m\sum_{\tau=t-D^{\max}}^{t-1} (\gamma^\tau_\ell)^2 \E{\sqnrm{\nabla_\ell\tilde{f}^\mu_\ell(\Theta^\tau_\ell)}}\\
        &+ L\frac{m D^{\max} + 1}{2} \frac{\tilde\sigma^2}{B} \sum_{\ell=1}^m (\gamma^t_\ell)^2\\
        &+ L\frac{m}{2} \frac{\tilde\sigma^2}{B} \sum_{\ell=1}^m\sum_{\tau=t-D^{\max}}^{t-1} (\gamma^\tau_\ell)^2.
    \end{align*}
    Let us telescope this inequality from $t=0$ until $t=\bar t$ to have
    \begin{align*}
        \E{{f}^{\mu}(\bm{\theta}^{\bar t+1})} &\leq {f}^{\mu}(\bm{\theta}^0)
        - \sum_{t=0}^{\bar t}\sum_{\ell=1}^m \gamma^t_\ell \E{\sqnrm{\nabla_\ell\tilde{f}^\mu_\ell(\Theta^t_\ell)}}\\
        &+ M \sum_{t=0}^{\bar t}\sum_{\ell=1}^m (\gamma^t_\ell)^2 \E{\sqnrm{\nabla_\ell\tilde{f}^\mu_\ell(\Theta^t_\ell)}}\\
        &+ M \frac{\tilde\sigma^2}{B} \sum_{t=0}^{\bar t}\sum_{\ell=1}^m (\gamma^t_\ell)^2,
    \end{align*}
    where $\sum_{t=0}^{\bar t} \sum_{\tau=t-D^{\max}}^{t-1} \{\cdot\} \leq D^{\max} \sum_{t=0}^{\bar t} \{\cdot\}$ is considered and 
    \[
        M \coloneqq L\frac{2m D^{\max} + 1}{2}
    \]
    is defined. Also, $\bm{\theta}^0$ is assumed to be deterministic due to initialization in \cref{alg:proposed}. Rearranging the above inequality
    \begin{align*}
        \eta \sum_{t=0}^{\bar t}  \bar\gamma^t \sum_{\ell=1}^m \E{\sqnrm{\nabla_\ell\tilde{f}^\mu_\ell(\Theta^t_\ell)}}
        &\leq {f}^{\mu}(\bm{\theta}^0) - {f}^{\mu^\star} \label{eq:convEnq} \numberthis\\
        &+ M \frac{\tilde\sigma^2}{B} \sum_{t=0}^{\bar t}\sum_{\ell=1}^m (\gamma^t_\ell)^2,
    \end{align*}
    where
    \[
        \eta \coloneqq 1 - M \gamma^{\max}, ~\gamma^{\max} \coloneqq \max_{i\in[m],t\in\N}[\gamma^t_i],
        ~\bar\gamma^t \coloneqq \min_{i\in[m]}[\gamma^t_i], ~ \forall t,  
    \]
    and $\E{{f}^{\mu}(\bm{\theta}^{\bar t+1})} \geq {f}^{\mu^\star}$ is considered where ${f}^{\mu^\star}$ minimizes ${f}^{\mu}$.
    In \eqref{eq:convEnq}, $\eta > 0$ guarantees a descent, hence $\gamma^{\max} < \nicefrac{1}{M}$ requires to hold. 
    Finally, we devide both sides of \eqref{eq:convEnq} by $\sum_{t=0}^{\bar t}  \bar\gamma^t$ and then lower bound the lhs as
    \begin{align*}
        \eta \sum_{t=0}^{\bar t}  \frac{\bar\gamma^t}{\sum_{k=0}^{\bar t}  \bar\gamma^k} &\sum_{\ell=1}^m \E{\sqnrm{\nabla_\ell\tilde{f}^\mu_\ell(\Theta^t_\ell)}}\\
        &\geq \eta \sum_{t=0}^{\bar t}  \frac{\bar\gamma^t}{\sum_{k=0}^{\bar t}  \bar\gamma^k} \E{\sqnrm{\nabla_i\tilde{f}^\mu_i(\Theta^t_i)}}\\
        &\geq \min_{t \in \{0,\cdots,\bar t\}} \eta\E{\sqnrm{\nabla_i\tilde{f}^\mu_i(\Theta^t_i)}}, \quad \forall i\in [m],
    \end{align*}
    where the weighted average is lowerbounded by the $\min\{\cdot\}$ operator in the last inequality.
    Hence, by \eqref{eq:convEnq}, $\forall i\in[m]$
    \begin{align*}
        \min_{t \in \{0,\cdots,\bar t\}} &\E{\sqnrm{\nabla_i\tilde{f}^\mu_i(\Theta^t_i)}} \\
        &\leq \sum_{t=0}^{\bar t}  \frac{\bar\gamma^t}{\sum_{k=0}^{\bar t}  \bar\gamma^k} \sum_{\ell=1}^m \E{\sqnrm{\nabla_\ell\tilde{f}^\mu_\ell(\Theta^t_\ell)}} \label{eq:conv_rate} \numberthis\\
        &\leq \frac{{f}^{\mu}(\bm{\theta}^0) - {f}^{\mu^\star} + M \frac{\tilde\sigma^2}{B} \sum_{t=0}^{\bar t}\sum_{\ell=1}^m (\gamma^t_\ell)^2}{\eta\sum_{k=0}^{\bar t}  \bar\gamma^k}.
    \end{align*}
    Consider
    \begin{align*}
        &\E{\sqnrm{\nabla_i\tilde{f}_i(\Theta^{t}_i)}}=\mathbb{E}{\sqnrm{\nabla_i\tilde{f}_i(\Theta^{t}_i)-\nabla_i\tilde{f}^\mu_i(\Theta^{t}_i)+\nabla_i\tilde{f}^\mu_i(\Theta^{t}_i)}} \label{eq:f_ft_relation}\\
        &\leq \frac{1}{2m} \mu^2 L^2(n+3)^{3} + 2\E{\sqnrm{\nabla_i\tilde{f}^\mu_i(\Theta^{t}_i)}},\numberthis
    \end{align*}
    due to \cref{prop:approximation:3} and the fact that $\tilde{f}_i$ is $\tilde L$-smooth with $\tilde L \leq \frac{1}{\sqrt{m}}L$ due to \cref{prop:lipschitz}. Also,
    \begin{equation*}
        |({f}^{\mu}(\bm{\theta}^0) - {f}^{\mu^\star}) - (f(\bm{\theta}^0)-f^\star)| \leq \mu^2Ln,
    \end{equation*}
    where \cref{prop:approximation:3} is invoked. 
    Substituting the above inequalities in \eqref{eq:conv_rate} completes the proof.
\end{appendixproof}

\bigskip
\begin{appendixproof}{thm:funval}

    \ref{thm:funval:1}: For $\ell \in [m]$
    \begin{talign*}
        &\E{\sqnrm{\bar\Theta^t_\ell - \Theta^t_\ell}} = \E{\sum_{i\in\mathcal{N}^{\ell}_{rx}} \sum_{k\in[m]} \sqnrm{\bm{\theta}^t_k - \bm\theta_k^{\mathcal{T}_{\ell}^{i,t}}}}\\
        &= \E{\sum_{i\in\mathcal{N}^{\ell}_{rx}} \sum_{k\in[m]} \sqnrm{\sum_{\tau=\mathcal{T}_{\ell}^{i,t}}^{t-1} \gamma^\tau_k \bm{\tilde{s}}_k^\tau } }\\
        &\leq D^{\max} \sum_{i\in\mathcal{N}^{\ell}_{rx}} \sum_{k\in[m]} \sum_{\tau=t-D^{\max}}^{t-1} (\gamma^\tau_k)^2 \E{\sqnrm{\bm{\tilde{s}}_k^\tau}} \label{eq:thm:funval}\numberthis\\
        & \leq m (D^{\max})^2 \frac{\tilde\sigma^2}{B} \sum_{k\in[m]}(\gamma^{t-D^{\max}}_k)^2 \\
        &+ m D^{\max} \sum_{k\in[m]} \sum_{\tau=t-D^{\max}}^{t-1} (\gamma^\tau_k)^2
        \E{\sqnrm{\nabla_k\tilde{f}^\mu_k(\Theta^\tau_k)}},
    \end{talign*}
    where in the first equality, the definitions of $\bar\Theta^t_\ell$ and $\Theta^t_\ell$ respectively in \eqref{eq:Params} and \eqref{eq:theta_common} are utilized, and in the first inequality, \cref{ass:asynchrony:2} and Jensen's inequality are invoked. 
    Moreover, in the last inequality, the bound in \eqref{eq:mean_s} and $|\mathcal{N}^{\ell}_{rx}|\leq m$ are used. 
    Also, since the stepsizes are diminishing, $\gamma^\tau_k \leq \gamma^{t-D^{\max}}_k, \forall \tau\in[t-D^{\max},t-1], \forall k\in[m]$ is considered.
    The last term can still be upperbounded further by
    \begin{align*}
        &\E{\sqnrm{\nabla_k\tilde{f}^\mu_k(\Theta^\tau_k)}} \\
        &= \E{\sqnrm{\nabla_k\tilde{f}^\mu_k(\Theta^\tau_k) - \nabla_k\tilde{f}_k(\Theta^\tau_k) + \nabla_k\tilde{f}_k(\Theta^\tau_k)}}\\
        &\leq \frac{1}{2m} \mu^2 L^2(n+3)^{3} + 2 \E{\sqnrm{\nabla_k\tilde{f}_k(\Theta^\tau_k)}}\\
        &\leq \frac{1}{2m} \mu^2 L^2(n+3)^{3} + 2 G^2 \eqqcolon \Delta
    \end{align*}
    where the first and second inequalities are due to \cref{prop:approximation:3}, with $\tilde L \leq \frac{1}{\sqrt{m}}L$ due to \cref{prop:lipschitz}, and \cref{ass:oracle:2}, respectively.
    Putting the last bound back into \eqref{eq:thm:funval} results in 
    \begin{talign*}
        \E{\sqnrm{\bar\Theta^t_\ell - \Theta^t_\ell}} &\leq m (D^{\max})^2 \frac{\tilde\sigma^2}{B} \sum_{k\in[m]}(\gamma^{t-D^{\max}}_k)^2 \label{eq:funval:1} \numberthis\\
        &+ m \Delta D^{\max} \sum_{k\in[m]} \sum_{\tau=t-D^{\max}}^{t-1} (\gamma^\tau_k)^2
    \end{talign*}
    Since $D^{\max}$ is finite due to \cref{ass:asynchrony} and \cref{prop:globalInfoMaxDel}, and the stepsizes are diminishing, the lhs vanishes when $t\to\infty$.

    \ref{thm:funval:3}: Using the Lipschitz continuity of $f_i$ guaranteed by \cref{ass:lipschitz:1}, 
    \begin{align*}
        \E{|f_i(\bm{\theta}^t) - f_i(\bm{\theta}^{\mathcal{T}_{\ell}^{i,t}})|} 
        \leq L^0 \E{\nrm{\bm{\theta}^t - \bm{\theta}^{\mathcal{T}_{\ell}^{i,t}}}} \to 0
    \end{align*}
    as $t\to\infty$, due to \ref{thm:funval:1}.

    \cref{thm:funval:2}: Take
    \begin{talign*}
        &\E{\sqnrm{\nabla f(\bm{\theta}^t)}} = \sum_{\ell=1}^m \E{\sqnrm{\nabla_\ell f(\bm{\theta}^t)}}\\ 
        &\leq 2\sum_{\ell=1}^m\E{\sqnrm{\nabla_\ell \tilde{f}_\ell(\bar\Theta^t_\ell)-\nabla_\ell \tilde{f}_\ell(\Theta^t_\ell)}} \\
        &+ 2\sum_{\ell=1}^m\E{\sqnrm{\nabla_\ell \tilde{f}_\ell(\Theta^t_\ell)}}\\
        &\leq 2 \tilde L^2 \sum_{\ell=1}^m \E{\sqnrm{\bar\Theta^t_\ell - \Theta^t_\ell}} + 2\sum_{\ell=1}^m\E{\sqnrm{\nabla_\ell \tilde{f}_\ell(\Theta^t_\ell)}},
    \end{talign*}
    where in the first inequality, Young's inequality and \eqref{eq:funcs:theta_common}, and in the second inequality, $\tilde L$-smoothness of $\tilde{f}_\ell$ established by \cref{prop:lipschitz} are considered.
    The bound is also valid if both sides are multiplied by $\alpha^t \coloneqq \nicefrac{\bar\gamma^t}{\sum_{k=0}^{\bar t} \bar\gamma^k}$, where $\bar\gamma^k$ is defined in \eqref{eq:convEnq}, and summed over $t$:
    \begin{talign*}
        &\sum_{t=0}^{\bar t} \alpha^t \E{\sqnrm{\nabla f(\bm{\theta}^t)}} \\
        &\leq 2 \tilde L^2 \sum_{t=0}^{\bar t} \alpha^t  \sum_{\ell=1}^m \E{\sqnrm{\bar\Theta^t_\ell - \Theta^t_\ell}} \\
        &+ 2\sum_{t=0}^{\bar t} \alpha^t \sum_{\ell=1}^m\E{\sqnrm{\nabla_\ell \tilde{f}_\ell(\Theta^t_\ell)}}. \label{eq:conv_inequality} \numberthis
    \end{talign*}
    With the defined stepsize sequence, the first term in the rhs is diminishing wrt $\bar t$ due to \eqref{eq:funval:1}. 
    Moreover, as the second term dominates the first term on the rhs, it can be concluded that the rhs diminishes at a rate of $\mathcal{O}(\mu^2) + \mathcal{O}(\nicefrac{1}{\sqrt{\bar t}})$, due to \eqref{eq:conv_rate} and \eqref{eq:f_ft_relation}.
    It is highlighted that inequalities \eqref{eq:conv_rate} and \eqref{eq:funval:1} are considered here with $\tilde\sigma^2$ defined in \cref{cor:twoPointQuery}, since the assumption that the oracles $F_\ell(\cdot,\bm{\xi}), \forall \ell\in[N]$, can be queried in any two points holds.
    The lhs in \eqref{eq:conv_inequality} can also be lowerbounded by the $\min\{\cdot\}$ operator, similar to \eqref{eq:conv_rate}, thus completing the proof.
\end{appendixproof}
% -------------------------------------------------------------------------
\bibliographystyle{IEEEtranTCOM}
\bibliography{IEEEabrv,TeX/bibliography/RA_bib.bib}

\end{document}